\documentclass[12pt]{article}

\usepackage{psfrag}
\usepackage[small]{caption}
\usepackage{color}
\definecolor{darkblue}{rgb}{0.1,0.1,.7}
\usepackage[dvips, colorlinks, linkcolor=darkblue, citecolor=darkblue, linktocpage]{hyperref}
\usepackage{epsfig,amsmath,amssymb,verbatim,mathrsfs}
\usepackage[square, comma, sort&compress,numbers]{natbib}
\usepackage{dsdshorthand}

\numberwithin{equation}{section}
\newcommand\cN{\mathcal{N}}
\newcommand\cJ{\mathcal{J}}
\newcommand\cV{\mathcal{V}}
\newcommand\cS{\mathcal{S}}
\newcommand\cU{\mathcal{U}}
\newcommand\cW{\mathcal{W}}
\newcommand\cK{\mathcal{K}}
\newcommand\tr{\mathrm{tr}}
\newcommand\prim{\mathrm{prim}}

\title{{\Large {\bf Bounds on 4D Conformal and Superconformal \\ Field Theories}}}
\author{David Poland and David Simmons-Duffin \\ \\
{\it \normalsize Jefferson Physical Laboratory, Harvard University,}\\
{\it \normalsize Cambridge, Massachusetts 02138, USA}}

\begin{document}

\begin{titlepage}

\noindent

\vspace{1cm}

\maketitle
\thispagestyle{empty}

\begin{abstract}
We derive general bounds on operator dimensions, central charges, and OPE coefficients in 4D conformal and $\mathcal N=1$ superconformal field theories. In any CFT containing a scalar primary $\f$ of dimension $d$ we show that crossing symmetry of $\<\f\f\f\f\>$ implies a completely general lower bound on the central charge $c\geq f_c(d)$.  Similarly, in CFTs containing a complex scalar charged under global symmetries, we bound a combination of symmetry current two-point function coefficients $\tau^{IJ}$ and flavor charges.  We extend these bounds to $\cN=1$ superconformal theories by deriving the superconformal block expansions for four-point functions of a chiral superfield $\Phi$ and its conjugate.  In this case we derive bounds on the OPE coefficients of scalar operators appearing in the $\Phi \times  \Phi^{\dagger}$ OPE, and show that there is an upper bound on the dimension of $\Phi^\dag\Phi$ when $\dim \Phi$ is close to $1$.   We also present even more stringent bounds on $c$ and $\tau^{I J}$.  In supersymmetric gauge theories believed to flow to superconformal fixed points one can use anomaly matching to explicitly check whether these bounds are satisfied.
\end{abstract}

\end{titlepage}

\setcounter{page}{1}

\setcounter{tocdepth}{2}
\tableofcontents

\vfill\eject


\newpage

\section{Introduction}
\label{sec:intro}

Near-conformal dynamics may describe physics beyond the Standard Model, in addition to facets of QCD itself.  Examples include walking~\cite{Holdom:1984sk,Akiba:1985rr,Appelquist:1986an,Yamawaki:1985zg,Appelquist:1986tr,Appelquist:1987fc} and conformal~\cite{Luty:2004ye,Luty:2008vs,Galloway:2010bp} technicolor, dynamical explanations of the flavor hierarchies~\cite{Georgi:1983mq, Nelson:2000sn, Poland:2009yb, Craig:2010ip}, solutions to the SUSY flavor problem~\cite{Kobayashi:2001kz, Nelson:2001mq, Luty:2001jh, Luty:2001zv, Kobayashi:2002iz, Dine:2004dv, Sundrum:2004un, Ibe:2005pj, Ibe:2005qv, Schmaltz:2006qs, Kachru:2007xp, Aharony:2010ch, Kobayashi:2010ye, Dudas:2010yh}, solutions to the $\mu/B\mu$ problem~\cite{Roy:2007nz, Murayama:2007ge, Perez:2008ng,Kim:2009sy,Craig:2009rk}, and so on.  While many of these ideas are promising, they often rely crucially on assumptions about the behavior of strongly-coupled field theories.  However, conformal symmetry itself severely restricts the structure of these theories, and it is not fully understood which assumptions are consistent with these restrictions and which are not.

In~\cite{Rattazzi:2008pe,Rychkov:2009ij,Caracciolo:2009bx} significant progress was made in understanding the range of behavior that is possible in 4D conformal field theories.  The key insight is that crossing symmetry of four-point functions requires that coefficients appearing in the operator product expansion (OPE) not be too large.  Combined with certain assumptions about the spectrum of operators, these constraints can potentially lead to a contradiction with unitarity, allowing one to rule out the spectrum.  Concretely, in~\cite{Rattazzi:2008pe,Rychkov:2009ij} it was shown that there is a completely general upper bound on the dimension of the lowest-dimension scalar primary operator appearing in the OPE $\f \times \f$ of a real scalar primary of dimension $d$ with itself, $\De_{\f^2} \leq f(d)$, where $f(d)$ is a function that is determined numerically.  In~\cite{Caracciolo:2009bx} it was also shown that one could compute an upper bound on the coefficient of the three-point function $\<\f\f\cO\>$ for any scalar primary $\cO$ appearing in the OPE.  

In the present work, we extend the analysis of~\cite{Rattazzi:2008pe,Rychkov:2009ij,Caracciolo:2009bx} in several directions.  First, we examine crossing symmetries of correlators involving charged fields in CFTs with global $U(1)$ symmetries, focusing in particular on the additional constraints that are present in superconformal theories.  We consider a chiral superconformal primary operator $\Phi$ of dimension $d$, and show how the four-point function $\<\Phi \Phi^{\dagger} \Phi \Phi^{\dagger}\>$ may be expanded in terms of ``superconformal blocks", which sum up the contributions of a given superconformal multiplet appearing in the $\Phi \times \Phi^{\dagger}$ OPE.  Since each superconformal multiplet contains a finite number of primary operators under the conformal sub-algebra, superconformal blocks may be decomposed into a finite sum of conformal blocks.  While such a decomposition was previously known in the context of $\cN=2$ and $\cN=4$ theories~\cite{Dolan:2001tt}, we believe that the $\cN=1$ result we present is new.  We further show how the $\cN=2$ superconformal blocks derived in~\cite{Dolan:2001tt} may be decomposed in terms of $\cN=1$ superconformal blocks, providing a non-trivial check on our result.

Second, we combine our superconformal block analysis with the methods of~\cite{Rattazzi:2008pe,Rychkov:2009ij} to derive bounds on the spectrum of operators appearing in the $\Phi\x\Phi^\dag$ OPE.  In particular, we find that there is an upper bound on the dimension of the $\Phi^\dag\Phi$ operator (defined as the lowest-dimension scalar appearing in $\Phi\x\Phi^\dag$) when $d$ is close to $1$.  Since the chiral operator $\Phi^2$ with dimension $2d$ always appears in $\Phi\x\Phi$, one cannot reproduce our bound on $\Phi^\dag\Phi$ by simply applying results from~\cite{Rattazzi:2008pe,Rychkov:2009ij} to the real or imaginary parts of $\Phi$.  We also compute bounds on the OPE coefficient of any scalar superconformal primary appearing in $\Phi\x\Phi^\dag$, independent of assumptions about the spectrum.  Our dimension and OPE bounds constitute completely general non-perturbative results about non-BPS quantities in $\cN=1$ superconformal theories.

Third, we use crossing relations among complex scalars to study OPEs involving conserved currents, both in the supersymmetric and non-supersymmetric context.  If $\f$ is a complex scalar primary of dimension $d$, then the OPE $\f\x\f^*$ contains global symmetry currents $J^{a I}$, whose coefficients are fixed by a Ward identity to be proportional to the charges of $\f$.  Crossing symmetry of $\<\f\f^*\f\f^*\>$ then implies an upper bound on the charges of $\f$ relative to the ``flavor central charges" $\tau^{I J}$, defined by the coefficient of the two-point function $\<J^{a I} J^{b J}\> \propto \tau^{IJ}$.  Specifically, we show $\tau_{I J} T^I T^J \leq f_{\tau}(d)$, where $\tau_{IJ}$ is the inverse of $\tau^{IJ}$, and $T^I$ are the global symmetry generators in the $\f$ representation.  We further strengthen this bound when $\f$ is the lowest component of a chiral multiplet $\Phi$ in a superconformal theory, in which case flavor currents appear as descendants of scalar operators $J^I$.  

Last, we turn to OPE's involving the stress tensor $T^{ab}$ in both supersymmetric and non-supersymmetric theories.  We show that in any CFT containing a real scalar primary operator $\f$ of dimension $d$, there is a completely general lower bound on the value of the central charge $c \geq f_c(d)$.  This again occurs because crossing symmetry of the four-point function $\<\f\f\f\f\>$ requires that the OPE coefficient in front of the stress tensor $\f(x)\f(0) \sim T^{a b}(0)$ not be too large, and this coefficient is fixed in terms of $c$ and $d$ by a Ward identity.  This can perhaps be viewed as a four-dimensional counterpart to the bound on $c$ derived in~\cite{Hellerman:2009bu} for two-dimensional CFTs.  Once again, we strengthen this bound in the supersymmetric case, where the stress tensor appears as a descendant of the $U(1)_R$ current in the $\Phi\x\Phi^\dag$ OPE.

Our bounds on $\tau_{I J} T^I T^J$ and $c$ are particularly interesting in supersymmetric theories since these quantities are determined in terms of the superconformal $U(1)_R$ symmetry as $\tau^{IJ} = -3 \Tr(R T^I T^J)$ and $c=\frac{1}{32}\left(9\Tr R^3-5 \Tr R\right)$, and may be calculated via 't Hooft anomaly matching.  Since $\Phi$ is chiral, its dimension is also determined in terms of the $U(1)_R$ symmetry as $d=\frac{3}{2} R$.  Thus, one may check whether these bounds are satisfied in the myriad asymptotically free $\cN=1$ theories that are believed to flow to superconformal fixed points, and we will demonstrate that this is the case in a few simple examples.  

\section{Preliminaries}
\label{sec:cft}
\subsection{CFT Review}
\label{sec:cftreview}
We will begin our discussion by reviewing some basic facts about 4D conformal field theories.  The conformal algebra may be written as
\ben
\,[M_{ab},P_c ] &=& P_a \eta_{bc} - P_b\eta_{ac}, \qquad [ M_{ab},K_c ] \,\,\,=\,\,\, K_a \eta_{bc} - K_b\eta_{ac}\nonumber\\
\,[M_{ab},M_{cd}] &=& \eta_{bc}M_{ad}-\eta_{ac}M_{bd}-\eta_{bd}M_{ac}+\eta_{ad}M_{bc}\nonumber\\
\,[D,P_a ] &=& P_a,\qquad [D,K_a]\,\,\,=\,\,\,-K_a\nonumber\\
\,[K_a,P_b] &=& 2\eta_{ab} D-2 M_{ab}, 
\een
and primary operators $\cO^I(0)$ are defined by the condition $K_{a}\cO^I(0)=0$,\footnote{For notational convenience we will leave the adjoint action of $K_a$ implicit in expressions such as this, so that $K_a \cO^I(0) \rightarrow [K_a, \cO^I(0)]$. Fermionic gradings should be respected.  For example if $\cO$ is bosonic, then $\bar Q^2 \cO$ is short for $\{\bar Q_{\dot\a},[\bar Q^{\dot\a},\cO]\}$.} where $K_a$ is the generator of special conformal transformations.  Fields may then be constructed by exponentiating the translation operator, $\cO^I(x) \equiv e^{x P} \cO^I(0)$.  Here $I$ denotes possible Lorentz indices, which can be labeled by $(j,\bar{j})$ according to the representation of $\SO(4) \cong \SU(2) \times \SU(2)$.  For example, traceless symmetric tensors $\cO^{a_1 \dots a_l}(x)$ have $j=\bar j=l/2$ (which we call the ``spin-$l$" representation).  We refer the reader to Appendix~\ref{app:conventions} for a more complete summary of the conventions used in this paper.

In 4D conformal field theories the correlation functions of primary operators are highly constrained (see e.g.~\cite{Osborn:1993cr}).  In particular, the two-point function for a spin-$l$  primary operator $\cO^{a_1 \dots a_l}(x)$ of dimension $\De$ can in general be written as
\ben\label{eq:OO2point}
\<\cO^{a_1 \dots a_l} (x_1) \cO^{b_1 \dots b_l}(x_2)\> &=& \frac{I^{a_1 b_1}(x_{12}) \dots I^{a_l b_l}(x_{12}) }{x_{12}^{2 \De}},\nonumber\\
I^{a b}(x) &\equiv& \eta^{a b} - 2 \frac{x^a x^b}{x^2},
\een
where $x_{12} \equiv x_1 - x_2$, and the indices $a_1 \dots a_l$ and $b_1 \dots b_l$ are implicitly symmetrized and made traceless.  Unitarity requires that the coefficient of the two-point function is positive, so that one can choose a basis of primary operators with the above normalization, where additionally two-point functions between different basis elements are taken to vanish, $\<\cO(x_1) \cO'(x_2)\> = 0$ for $\cO \neq \cO'$.  Positivity of the two-point functions of {\it descendant} operators then further imposes the unitarity bounds~\cite{Mack:1975je}
\ben
\De &\geq& 1 \qquad\,\,\,\,\,\,\,\,\, (l=0), \nonumber\\
\De &\geq& l+2  \qquad (l \geq 1).
\een

Three-point functions between scalar primary operators $\phi_i(x)$ of equal dimension $d$ and a spin-$l$ primary $\cO^{a_1 \dots a_l}(x)$ of dimension $\De$ are fixed up to an overall constant as
\ben
\<\phi_{1}(x_1)\phi_{2}(x_2) \cO^{a_1\dots a_l}(x_3) \> &=& \frac{ \lambda_{\phi_{1}\phi_{2}\cO}}{x_{12}^{2d-\De+l} x_{23}^{\De-l} x_{13}^{\De-l}} Z^{a_1}\dots Z^{a_l},\nonumber\\
Z^a &\equiv& \frac{x_{31}^a}{x_{31}^2} - \frac{x_{32}^a}{x_{32}^2}.
\een
If we take $\phi_1 = \phi_2$, it is straightforward to see that invariance under $x_1 \leftrightarrow x_2$ requires that $l$ must be even in order for the three-point function to be non-vanishing.  However, if $\phi_1 \neq \phi_2$ then odd-$l$ primaries are also allowed.  Note that Lorentz representations with $j\neq\bar{j}$ cannot appear because there does not exist a function built out of the $x_i$'s having the required transformation properties.  As reviewed in~\cite{Rattazzi:2008pe}, when $\f_1$ and $\f_2$ are real, the coefficients $\l_{\phi_1\phi_2\cO}$ are necessarily real in the basis of Eq.~(\ref{eq:OO2point}).

Finally, four-point functions of scalar operators are not completely determined by symmetry considerations alone, and in the case of equal dimensions can always be written as
\ben
\label{eq:general4ptfn}
\<\phi_{1}(x_1)\phi_{2}(x_2)\phi_{3}(x_3)\phi_{4}(x_4)\> &=& \frac{g(u,v)}{x_{12}^{2d} x_{34}^{2d}},
\een
where $g(u,v)$ is a function of the conformally-invariant cross ratios $u\equiv \frac{x_{12}^2 x_{34}^2}{x_{13}^2 x_{24}^2}$ and $v \equiv \frac{x_{14}^2 x_{23}^2}{x_{13}^2 x_{24}^2}$.  Though $g(u,v)$ is not fixed by conformal symmetry, it is fully determined by the dynamical data of the theory, namely the spectrum of operator dimensions and spins $\De,l$ and three-point function coefficients $\l_{\f_1\f_2\cO}$. This is most easily seen through the operator product expansion (OPE), which relates a product of operators at different positions to a sum over operators at a single position.  In the case of scalar primaries, we can write
\ben
\phi_1(x) \phi_2(0) &=& \sum_{\cO \in \phi_1 \times \phi_2} C_I(x,P) \cO^I(0),
\een
where $I$ stands for possible Lorentz indices. We use the notation $\cO\in\f_1\x\f_2$ to mean that the sum should be taken over {\it primary} operators occurring in the OPE of $\f_1$ with $\f_2$.  The operator $C_I(x,P)$ may for example be determined by inserting the OPE into the three-point functions and using the known form of the two-point functions~\cite{Dolan:2000ut}.  

Taking the $\phi_1(x_1) \times \phi_2(x_2)$ OPE and the $\phi_3(x_3)\times \phi_4(x_4)$ OPE in the four-point function then leads to the conformal block decomposition
\ben
g(u,v) &=& \sum_{\cO \in \phi_1 \times \phi_2} \l_{\phi_1\phi_2\cO} \l_{\phi_3 \phi_4 \cO} g_{\De,l}(u,v) ,
\een
where the ``conformal blocks" $g_{\De,l}(u,v)$ are given explicitly by~\cite{Dolan:2000ut}
\ben
\label{eq:explicitconformalblocks}
g_{\De,l}(u,v) &=& \frac{(-1)^l}{2^l} \frac{z \bar{z}}{z-\bar{z}} \left[k_{\De+l}(z) k_{\De-l-2}(\bar{z}) - z \leftrightarrow \bar{z}\right] \nonumber\\
k_{\b}(x) &=& x^{\b/2} {}_2 F_1(\b/2,\b/2,\b;x),
\een
and the change of variables $u = z \bar{z}$ and $v=(1-z)(1-\bar{z})$ has been used.
We note in passing that the conformal blocks can also be elegantly derived by viewing them as eigenfunctions of the quadratic casimir of the conformal group~\cite{Dolan:2003hv}.

If we take all of the scalars to be identical, then invariance of Eq.~(\ref{eq:general4ptfn}) under $x_1 \leftrightarrow x_3$ leads to the ``crossing symmetry" constraint
\ben\label{eq:crossing}
\sum_{\cO \in \phi \times \phi} \l_{\phi\phi\cO}^2 g_{\De,l}(u,v) &=& \left(\frac{u}{v}\right)^{d} \sum_{\cO \in \phi \times \phi} \l_{\phi\phi\cO}^2 g_{\De,l}(v,u),
\een
which must satisfied by any consistent spectrum of dimensions, spins, and choice of three-point function coefficients.  A key point is that unitarity requires $\l_{\f\f\cO}\in \R$, so the coefficients $\l^2_{\f\f\cO}$ appearing above are positive.  Invariance under $x_1 \leftrightarrow x_2$  again tells us that only even-spin operators may appear, and other exchanges do not give any new information.

\subsection{Bounds from Crossing Relations}
\label{sec:boundsfromcrossing}

In~\cite{Rattazzi:2008pe,Rychkov:2009ij}, the crossing relation of Eq.~(\ref{eq:crossing}) was used to derive an upper bound on the dimension of the lowest-dimension scalar operator appearing in the OPE $\f\x\f$.  In~\cite{Caracciolo:2009bx} bounds were also derived on the size of the three-point function coefficients of scalar operators appearing in $\f\x\f$.  The techniques employed depend on the explicit expression Eq.~(\ref{eq:explicitconformalblocks}) for conformal blocks, together with the unitarity requirement $\l_{\f\f\cO}^2\geq 0$.  We now review these techniques; in the following subsection we will discuss some generalizations.

\begin{figure}[h!]
\begin{center}
\begin{psfrags}
\psfrag{S}[B][B][1.2][0]{$\mathcal{S}$}
\psfrag{H}[l][B][1][0]{$\a(F_{\De_0,l_0})=1$}
\psfrag{V}[B][B][1.2][0]{$\mathcal{V}^*$}
\psfrag{F}[B][B][1][0]{$\a(F_{\De,l})\geq0$}
\includegraphics[width=80mm]{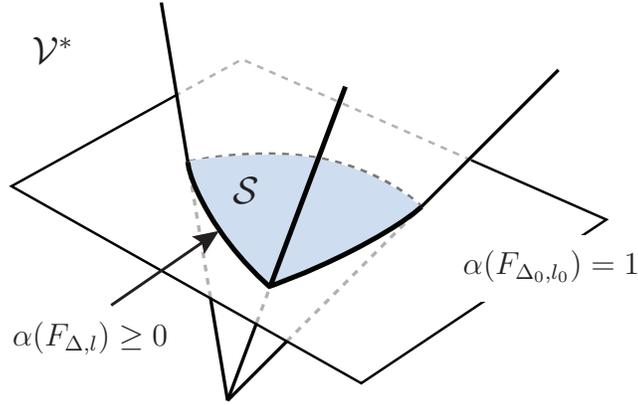}
\end{psfrags}
\end{center}
\caption{The ``search space" $\cS\subset \cV^*$ is the intersection of the hyperplane $\a(F_{\De_0,l_0})=1$ with the convex cone of linear functionals $\a$ satisfying $\a(F_{\De,l})\geq 0$ for all $(\De,l)$ in the spectrum.}
\label{fig:searchspace}
\end{figure}

Let us begin by showing how to bound the OPE coefficient-squared $\l_{\cO_0}^2 \equiv \l_{\f\f\cO_0}^2$ of a given operator $\cO_0$ of dimension $\De_0$ and spin $l_0$ appearing in $\f\x\f$.  We first rewrite the crossing relation by separating out and dividing by the contribution of the unit operator, as well as separating out the contribution of the particular operator $\cO_0$ whose OPE coefficient we would like to study,
\ben
\label{eq:rewritecrossing}
\l_{\cO_0}^2 F_{\De_0,l_0}(u,v) &=& 1 - \sum_{\cO\neq \cO_0}\l_\cO^2 F_{\De,l}(u,v),
\een
where 
\ben
\label{eq:FdeltaL}
F_{\De,l}(u,v) &\equiv& \frac{v^d g_{\De,l}(u,v) - u^d g_{\De,l}(v,u)}{u^d-v^d},
\een
and we have used that $g_{0,0}(u,v)=1$ for the unit operator.  Note that $F_{\De,l}$ depends on $d$, though we are suppressing this dependence for brevity.  Eq.~(\ref{eq:rewritecrossing}) is a linear equation in the space $\cV$ of functions of two variables which are invariant under $u\leftrightarrow v$.  It encodes an infinite number of relations between OPE coefficients $\l_\cO$, but general statements about solutions consistent with unitarity ($\l_\cO^2 \geq 0$) can be difficult to extract.  The approach of \cite{Rattazzi:2008pe,Rychkov:2009ij,Caracciolo:2009bx} is to consider a real linear functional $\a\in \cV^*=\Hom(\cV,\R)$, which satisfies 
\ben
\label{eq:alphaconstraint1}
\a(F_{\De_0,l_0})&=&1,\quad\textrm{ and }\\
\label{eq:alphaconstraint2}
\a(F_{\De,l}) &\geq& 0,\quad\textrm{ for all other operators in the spectrum.}
\een
Then applying $\a$ to both sides of Eq.~(\ref{eq:rewritecrossing}), we obtain a bound
\ben
\l^2_{\cO_0}\ \ =\ \ \a(1)-\sum_{\cO\neq \cO_0}\l_{\cO}^2\a(F_{\De,l})\ \ \leq\ \ \a(1),
\een
where we have used that $\l_\cO^2\geq 0$ by unitarity.  Let us denote by $\cS$ the subspace of $\a\in \cV^*$ which satisfy the constraints (\ref{eq:alphaconstraint1}, \ref{eq:alphaconstraint2}) (depicted in Figure~\ref{fig:searchspace}). In many cases of interest, $\cS$ is non-empty, so a non-trivial bound on the OPE coefficient-squared $\l_{\cO_0}^2$ exists.

Given bounds on $\l^2_{\cO_0}$, bounds on the dimension of $\cO_0$ may or may not follow as a consequence.  For example, suppose we assume that $\cO_0$ with dimension $\De_0$ is the lowest-dimension scalar appearing in $\f\x\f$.  Then if we can find some $\a$ such that $\l_{\cO_0}^2 \leq \a(1)<0$, then we have found a contradiction with unitarity, implying that it's impossible that $\cO_0$ has dimension $\De_0$.

Now our bound $\l_{\cO_0}^2\leq \a(1)$ is most interesting when $\a(1)$ is as small as possible.  Thus we would like to minimize $\a(1)$ over all $\a\in \cS$.  This problem resembles an infinite-dimensional version of a {\it linear program}, which usually refers to a linear optimization problem over $\R^n$, subject to a finite number of affine constraints.  Linear programs have been well-studied in mathematics and computer science, and a number of efficient algorithms for their solution are known.  A key observation is that since the search space is an intersection of half-spaces (one for each inequality) and hyperplanes (one for each equality), it is convex.  Consequently, the optimum of any linear function lies on the boundary of the search space, and can be reached deterministically by following the direction of steepest descent (either along the boundary or in the interior).

A first step towards making our problem tractable via these methods is to restrict to a finite-dimensional subspace $\cW\subset \cV^*$.  Then, minimizing $\a(1)$ over $\a\in \cW\cap \cS$ will give a possibly sub-optimal, but still valid bound $\l_{\cO_0}^2\leq \a(1)$.  The choice of $\cW$ is somewhat arbitrary and unfortunately can have a significant effect on the answer.  A convenient class of subspaces is given by taking linear combinations of derivatives at some point in $z,\bar z$ space.  Following \cite{Rattazzi:2008pe,Rychkov:2009ij,Caracciolo:2009bx}, we take these derivatives around the point $z=\bar z=1/2$ (which is invariant under $u\leftrightarrow v$).  That is, we define $\cW_k\subset \cV^*$ to be the space of functionals
\ben
\label{eq:Wksuspaces}
\a : F(z,\bar{z}) &\mto& \sum_{m+n\leq 2k} a_{mn}\ptl_z^m\ptl_{\bar z}^nF(1/2,1/2)
\een
with real coefficients $a_{mn}$.\footnote{In addition to being simple to describe, the spaces $\cW_k$ are computationally convenient, since one can in fact derive relatively simple analytic expressions for derivatives of the functions $F_{\De,l}(z,\bar z)$ at $z=\bar z=1/2$, and these expressions can be computed efficiently using recursion relations (see Appendix \ref{app:implementation}).}  We can then scan over $\cW_k\cap \cS$ by varying the $a_{mn}$, subject to the constraints of Eqs.~(\ref{eq:alphaconstraint1}, \ref{eq:alphaconstraint2}).  One hopes that as we take $k\to \oo$, our search will cover more and more of $\cS$, and our bound will converge to the optimal one.\footnote{Here, ``optimal" means ``optimal given our assumptions," namely the diagonal crossing relation Eq.~(\ref{eq:crossing}) and unitarity of each OPE coefficient in $\f\x\f$.  These are a small subset of the full consistency relations of a CFT, so it's certainly possible that inputting more information could lead to even stronger bounds.}

Even after restricting to $\cW_k$, our problem differs from a typical linear program in that Eq.~(\ref{eq:alphaconstraint2}) includes an infinite number of affine constraints on $\a$.  For example, if we are interested in bounding $\l_{\cO_0}^2$ with no additional assumptions on the spectrum, then we must demand $\a(F_{\De,l})\geq 0$ for all $(\De,l)$ obeying the unitarity bound.  Alternatively, if we wish to bound the OPE coefficient of the lowest-dimension scalar in $\f\x\f$, we must take $\a(F_{\De,l})\geq 0$ for all scalars with $\De\geq \De_0$, and all $(\De,l)$ with $l>0$ that obey unitarity.  In each case, we have a continuously infinite number of constraints on $\a$ --- one for each $(\De,l)$ pair.

In a typical linear program, the search space is a convex polytope in $\R^n$, given by an intersection of a finite number of half-spaces and hyperplanes.  In our case, the search space $\cS\subset \cV^*$ is still convex, since it is an intersection of half-spaces $\cU_{\De,l}=\{\a: \a(F_{\De,l})\geq 0\}$ and a hyperplane $\cH=\{\a:\a(F_{\De_0,l_0})=1\}$.  However, because $\De$ can vary continuously, $\cS$ is not a polytope.  In general, the intersection $\cS\cap \cW$ with any finite-dimensional $\cW$ is ``piecewise-curved," e.g. it has (not-necessarily flat) faces, (possibly curved) edges, vertices, etc.  Consequently, we expect that at finite $k$, as we vary the underlying parameters of our problem ($d$, $\De_0$, etc.), our bound will vary in a ``piecewise-curved" way, with corners as the optimal $\a=\a_*$ passes over edges on the boundary of $\cS\cap \cW_k$.

In order to apply linear programming techniques, we need to approximate $\cS\cap \cW_k$ by a polytope.\footnote{Actually, there do exist algorithms to solve more general classes of ``convex optimization problems", which can involve curved search spaces.  It might be interesting to investigate whether any of these can be applied to crossing relations.}  Let us pick some finite discrete set $D=\{(\De_i,l_i)\}$ and reduce the constraints in Eq.~(\ref{eq:alphaconstraint2}) to simply $\a(F_{\De_i,l_i})\geq 0$ for all $(\De_i,l_i)\in D$.  This {\it expands} the search space, and we are now in danger of obtaining an invalid bound if the optimal $\a=\a_*$ satisfies $\a_*(F_{\De',l'})<0$ for some $(\De',l')$ not in $D$.  However, this danger disappears as we increase the size of $D$ and approximate $\cS\cap\cW_k$ by more and more refined polytopes.  A type of discretization that works well in practice is 
\ben
\label{eq:discretization}
D &=& \{(\De_\mathrm{min}+n\e,l):n=0,\dots,N\textrm{ and }l=0,2,\dots,L\},
\een
where $N\e$ and $L$ are large numbers (say $\sim 50$), and $\e$ is some small step size (say $\e\sim .1$ or $.01$).  By decreasing $\e$, we can ensure that violations of our constraints $\a_*(F_{\De',l'})<0$ become less and less important.  One must also ensure that $\a(F_{\De,l})$ is greater than zero asymptotically as $\De,l\to \oo$.  This is easy to check using the analytic expressions for derivatives of $F_{\De,l}$ given in Appendix \ref{app:implementation}.  In practice, the optimal $\a=\a_*$ often obeys the asymptotic constraint automatically, provided $N\e$ and $L$ are sufficiently large.

\subsubsection{Solutions to the Crossing Relations from Linear Programs}

In this subsection, we will show how in principle, the linear program described above produces not just an OPE coefficient bound, but also the corresponding ``optimal" solution to the crossing relations consistent with the given assumptions.  This type of solution doesn't necessarily have anything to do with CFTs, since we're only inputting a subset of the full CFT consistency relations.  However, we mention it here because it helps give some intuition for properties of optimal solutions $\a_*\in \cS$.  The results of this subsection are not used elsewhere in the paper, so the reader should feel free to skip to Section~\ref{sec:limitationsofrattazzi} if desired.

Let us briefly introduce some notation.  A subset $K\subset V$ of a finite-dimensional real vector space $V$ is called a {\it convex cone} if $\l_1 x_1+\l_2 x_2\in K$ for all $x_1,x_2\in K$ and $\l_1,\l_2\in \R_+$.  The {\it dual cone} of $K$ is the space of linear functions $K^\vee=\{\ell\in V^*\textrm{ such that }\ell(x) \geq 0\textrm{ for all }x\in K\}$.  One can show that if $K$ is closed, then $(K^\vee)^\vee=K$.

In the following, let us be cavalier and pretend that the space $\cV^*$ of possible $\a$'s is finite-dimensional.  Suppose we have run our linear program and arrived at an optimal $\a_*\in \cS$ that minimizes $\a(1)$ subject to Eqs.~(\ref{eq:alphaconstraint1}) and (\ref{eq:alphaconstraint2}).  Since $\a_*$ lies on the boundary of $\cS$, some of the constraints defining $\cS$ must be saturated at $\a_*$.  That is, there is some set $\{F_{\De_i,l_i}\textrm{ for }i=1,2,\dots\}$ such that $\a_*(F_{\De_i,l_i})=0$.  We also have $\a_*(F_{\De_0,l_0})=1$ by assumption.

For the local geometry of $\cS$ near $\a_*$, only the constraints which are saturated at $\a_*$ are important.  In other words, we can imagine locally replacing $\cS$ with the set $\a_*+\cK_{\a_*}$, where $\cK_{\a_*}$ is the convex cone
\ben
\cK_{\a_*} &\equiv& \{\b: \b(F_{\De_i,l_i})\geq 0\textrm{ for $i\geq 1$, and }\b(F_{\De_0,l_0})=0\} .
\een
Note that $\cK_{\a_*}$ is the dual cone of
\ben
\cK_{\a_*}^\vee &=& \R F_{\De_0,l_0}+\sum_{i\geq 1} \R_+ F_{\De_i,l_i},
\een
namely the positive span of the $F_{\De_i,l_i}$, plus $F_{\De_0,l_0}$ with an arbitrary real coefficient.

Now, the condition that $\a_*$ minimize $\a_*(1)$ means that $(\a_*+\de\a)(1)\geq \a_*(1)$ for all $\de\a$ pointing into the interior of $\cS$, that is all $\de\a\in \cK_{\a_*}$.  But this just means that $1$ is an element of the dual cone $\cK_{\a_*}^\vee$, so that there exist coefficients $q\in \R$ and $p_i \in \R_+$ with
\ben
\label{eq:optimalsolution}
1 &=& q F_{\De_0,l_0} + \sum_{i\geq 1} p_i F_{\De_i,l_i} .
\een
In other words, the $(\De_i,l_i)$ whose constraints are saturated at $\a_*$, along with $(\De_0,l_0)$, give the spectrum of a solution to the crossing relation Eq.~(\ref{eq:rewritecrossing}).  Further, if $q>0$ then this solution is consistent with unitarity, since the $p_i$ are positive.

\begin{figure}[h!]
\begin{center}
\begin{psfrags}
\psfrag{S}[B][B][1.2][0]{$\mathcal{S}$}
\psfrag{a}[B][B][1][0]{$\a_*$}
\psfrag{F}[B][B][1][0]{$-F_{\De_1,l_1}$}
\psfrag{G}[l][B][1][0]{$-F_{\De_2,l_2}$}
\psfrag{N}[B][B][1][0]{$-1$}
\includegraphics[width=50mm]{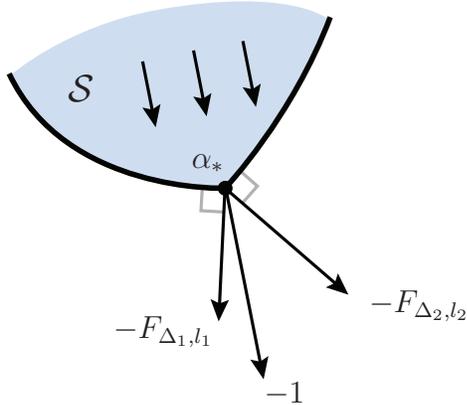}
\end{psfrags}
\end{center}
\caption{Picking a metric on $\cV^*$, we arrive at the following picture. The linear functional that minimizes $\a(1)$ is the unique point $\a_*$ on the boundary of $\cS$ where $-1$ is in the positive span of the ``normal vectors" $-F_{\De_i,l_i}$ to $\cS$ at $\a_*$.  Here, the three parallel arrows illustrate the direction of steepest descent of $\a(1)$.  We have suppressed an infinite number of dimensions (including the $F_{\De_0,l_0}$ direction) in order to draw this figure in the plane.}
\label{fig:positivespan}
\end{figure}

A useful geometric picture (illustrated in Figure~\ref{fig:positivespan}) for arriving at the above result is to imagine picking a metric and thinking of $-F_{\De_i,l_i}$ as specifying ``normal vectors" to the search space $\cS$ at $\a_*$.  The minimum of $\a(1)$ occurs precisely when the vector $-1\in \cV$ (which points in the direction we want to go) is in the positive span of normal vectors to $\cS$.    Meanwhile, since we can never move off the hyperplane $\a(F_{\De_0,l_0})=1$, it doesn't matter whether $-1$ has a component in the direction of $F_{\De_0,l_0}$, which is why $q$ can have either sign.

\begin{figure}
\begin{center}
\begin{psfrags}
\def\PFGstripminus-#1{#1}%
\def\PFGshift(#1,#2)#3{\raisebox{#2}[\height][\depth]{\hbox{%
  \ifdim#1<0pt\kern#1 #3\kern\PFGstripminus#1\else\kern#1 #3\kern-#1\fi}}}%
\providecommand{\PFGstyle}{}%
%
\psfrag{aF}[bc][bc]{\PFGstyle $\a_*(F_{\De,l})$}%
\psfrag{d01}[cc][cc]{\PFGstyle $\ \ (\De_0,1)$}%
\psfrag{d}[cl][cl]{\PFGstyle $\De$}%
\psfrag{l0}[cc][cc]{\PFGstyle $l\!=\!0$}%
\psfrag{l2}[cc][cc]{\PFGstyle $l\!=\!2$}%
\psfrag{l4}[cc][cc]{\PFGstyle $l\!=\!4$}%
\psfrag{x0}[tc][tc]{\PFGstyle $0$}%
\psfrag{x12}[tc][tc]{\PFGstyle $10$}%
\psfrag{x21}[tc][tc]{\PFGstyle $2$}%
\psfrag{x41}[tc][tc]{\PFGstyle $4$}%
\psfrag{x61}[tc][tc]{\PFGstyle $6$}%
\psfrag{x81}[tc][tc]{\PFGstyle $8$}%
\psfrag{y0}[cr][cr]{\PFGstyle $0$}%
\psfrag{y21}[cr][cr]{\PFGstyle $2$}%
\psfrag{y41}[cr][cr]{\PFGstyle $4$}%
\psfrag{y61}[cr][cr]{\PFGstyle $6$}%
\psfrag{y81}[cr][cr]{\PFGstyle $8$}%
\psfrag{ym21}[cr][cr]{\PFGstyle $-2$}%
\includegraphics{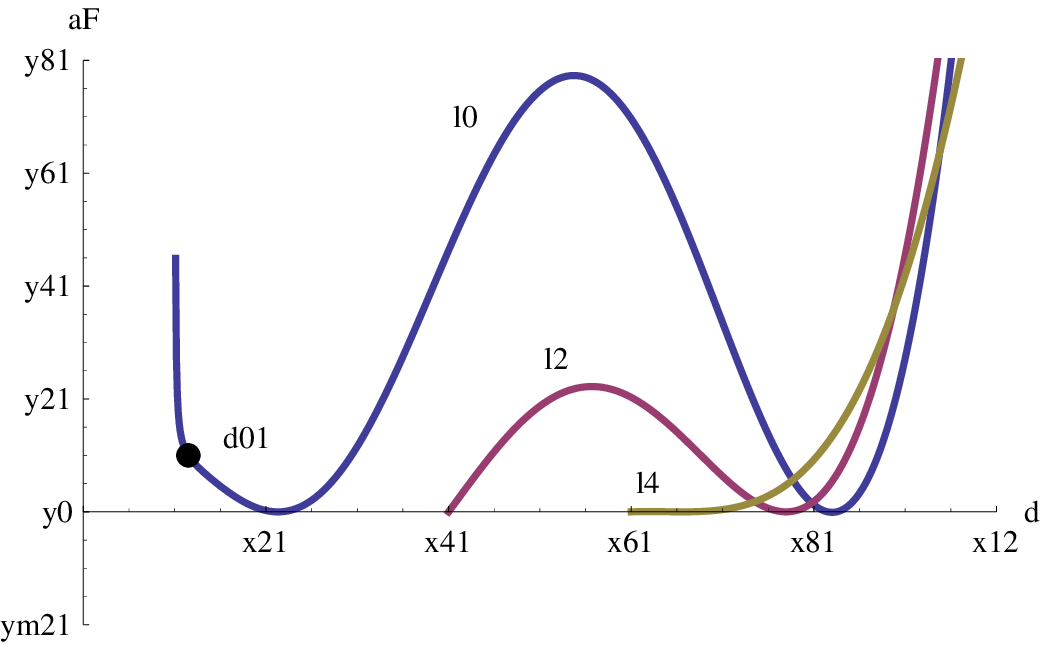}
\end{psfrags}
\end{center}
\caption{A plot of $\a_*(F_{\De,l})$ for various $\De,l$, where $\a_*\in\cS$ gives the strongest bound on the OPE coefficient of the lowest-dimension scalar $\cO_0\in\f\x\f$.  Here, we have taken $\dim\f=1.1$, $\De_0=\dim\cO_0=1.15$, and $k=4$.  We show only pairs $\De,l$ satisfying unitarity.  Note that $\a_*(F_{\De,l})$ is never negative in this range, consistent with the constraints of our linear program, although it has zeros $(\De,l)\in\{(2.2,0),(8.2,0),(4,2),(7.7,2),(6,4)\}$.  Note also that $\a_*(F_{\De_0,0})=1$, as required.}
\label{fig:functionalzeros}
\end{figure}

In practice, we must solve our linear program by first restricting to a finite dimensional search space $\cS\cap \cW_k$.  In this case, the optimal $\a_*$ will have a few saturated constraints $\a_*(F_{\De_i,l_i})=0$ ($i=1,\dots,N_k$).  We can see this explicitly in Figure~\ref{fig:functionalzeros} which plots $\a_*(F_{\De,l})$ for varying $\De$ and $l$, where $\a_*$ is the solution to a linear program.  Note that $\a_*(F_{\De,l})$ has zeros at particular $(\De,l)$, but of course never becomes negative.  As we increase $k$, we expect new zeros of $\a_*(F_{\De,l})$ to appear, with $N_k$ eventually running off to infinity.  If the limit $k\to\oo$ is ``well-behaved" in some appropriate sense, we might hope that the zeros $(\De_i,l_i)$ for small $\De$ and $l$ values converge quickly as $k\to\oo$, giving us some information about the low-dimension and spin part of the spectrum corresponding to the ``optimal" solution Eq.~(\ref{eq:optimalsolution}).  Indeed, this seems to be the case in practice.  It would be interesting to see if this information has any practical applications.

\subsection{Limitations and Generalizations}
\label{sec:limitationsofrattazzi}

One limitation of the formalism outlined above and the one used in~\cite{Rattazzi:2008pe,Rychkov:2009ij,Caracciolo:2009bx} is that one only learns about the OPE of a real scalar with itself.  In particular, the formalism does not allow one to distinguish between operators appearing in the OPE that have different global symmetry charges.  For example, in $\cN=1$ superconformal field theories there is a global $U(1)_R$ symmetry, and chiral operators have dimension $d  = \frac{3}{2} R$.  If we take $\phi$ to be the lowest component of a chiral multiplet $\Phi$, then the $\Re[\phi] \times \Re[\phi]$ OPE will contain both operators in the $\phi \times \phi$ OPE having $U(1)_R$ charge $2 R_{\Phi}$, and operators in the $\phi \times \phi^{*}$ OPE that are neutral under $U(1)_R$.  Since the $\phi^2$ operator appearing in the $\phi \times \phi$ OPE is chiral, it always has dimension $2d$ and automatically satisfies the bounds derived in~\cite{Rattazzi:2008pe,Rychkov:2009ij}.  Thus, we unfortunately do not learn anything new about the $U(1)_R$-singlet non-chiral operators appearing in $\phi \times \phi^{*}$.\footnote{Another example, extensively discussed in~\cite{Rattazzi:2008pe}, is that the bounds do not distinguish between $SU(2)$-singlet and $SU(2)$-triplet operators appearing in the $h \times h$ OPE in conformal technicolor scenarios.}

With this motivation in mind, let us consider more carefully what crossing relations apply in the case of a complex scalar charged under a global $U(1)$ symmetry.   We must first determine the unitarity constraints for three-point functions involving a complex scalar.  We know that the correlator with a real spin-$l$ operator $\<\phi(x_1) \phi^*(x_2) \cO^{a_1 \dots a_l}(x_3)\>$ must be invariant under the exchange $x_1 \leftrightarrow x_2$ combined with complex conjugation.  From this we learn that even-$l$ operators must have coefficients that are real, $\l_{\phi\phi^*\cO} = \l_{\phi\phi^*\cO}^*$, and odd-$l$ operators must have coefficients that are imaginary, $\l_{\phi\phi^*\cO} = -\l_{\phi\phi^*\cO}^*$.  On the other hand, the three-point function $\<\phi(x_1) \phi(x_2) \cO^{a_1 \dots a_l *}(x_3)\>$ must simply be invariant under $x_1 \leftrightarrow x_2$, and hence only even-$l$ operators may appear.  In this case, however, the coefficient $\l_{\phi\phi\cO^*}$ is in general complex.  Of course, the above arguments reproduce exactly what we would have concluded by breaking $\f$ into its real and imaginary parts $\f=\f_1+i\f_2$, and using the requirement from Section~\ref{sec:cftreview} that $\l_{\f_i\f_j\cO}\in\R$.

Now let us consider the four-point function $\<\phi(x_1)\phi(x_2)\phi^*(x_3)\phi^*(x_4)\>$.  We can evaluate this in two qualitatively different ways: by taking the $\phi(x_1)\x\phi(x_2)$ and $\phi^*(x_3)\times\phi^*(x_4)$ OPEs, or alternatively by taking $\f(x_1)\x\f^*(x_4)$ and $\phi(x_2) \times \phi^*(x_3)$.  Equating the resulting expressions leads to the crossing relation
\ben\label{eq:crossingphiphi}
\sum_{\cO \in \phi \times \phi} |\l_{\phi\phi \cO^*}|^2 g_{\De,l}(u,v) &=& \left(\frac{u}{v}\right)^{d} \sum_{\cO \in \phi \times \phi^*} |\l_{\phi\phi^*\cO}|^2 g_{\De,l}(v,u).
\een
Although one could conceivably apply the ideas of Section~\ref{sec:boundsfromcrossing} to this kind of relation, we have had more success applying linear programs to crossing relations that display symmetry under $u\leftrightarrow v$ --- that is, relations which involve the same spectrum of operators on both sides.  As we will see, Eq.~(\ref{eq:crossingphiphi}) implies two such independent relations that must be satisfied in a consistent theory.  By adding the equation to itself we can immediately derive one of them,
\ben\label{eq:crossingall}
\sum_{\substack{ \cO \in \phi \times \phi\phantom{*} \\  \cO \in \phi \times \phi^*}} |\l_{\cO}|^2 g_{\De,l}(u,v)  &=& \left(\frac{u}{v}\right)^{d} \sum_{\substack{ \cO \in \phi \times \phi\phantom{*} \\  \cO \in \phi \times \phi^*}} |\l_{\cO}|^2 g_{\De,l}(v,u),
\een
where $\l_{\cO}$ is shorthand for the appropriate three-point function coefficient.

The simplest way to see the second crossing symmetry constraint is to alternatively relabel the coordinates and consider expanding the four-point function $\<\phi(x_1) \phi^*(x_2) \phi(x_3) \phi^*(x_4)\>$ by taking the $\phi(x_1) \times \phi^*(x_2)$ OPE and the $\phi(x_3) \times \phi^*(x_4)$ OPE.  Exchanging $x_1 \leftrightarrow x_3$ then leads to the constraint
\ben\label{eq:crossingphiphis}
\sum_{\cO \in \phi \times \phi^*} |\l_{\phi\phi^*\cO}|^2 (-1)^l g_{\De,l}(u,v) &=& \left(\frac{u}{v}\right)^{d} \sum_{\cO \in \phi \times \phi^*} |\l_{\phi\phi^*\cO}|^2 (-1)^l g_{\De,l}(v,u),
\een
which is a crossing symmetry relation that only involves operators in the $\phi \times \phi^*$ OPE.  Here the $(-1)^l$ factors appear because the odd-$l$ operators necessarily have 3-point function coefficients that are imaginary, and their square (and not absolute value squared) enters the conformal block decomposition when the $\phi$'s have the above ordering.  One can show that Eq.~(\ref{eq:crossingphiphis}) in fact follows from Eq.~(\ref{eq:crossingphiphi}) through repeated use of the identity $g_{\De,l}(u,v) = (-1)^l g_{\De,l}(u/v,1/v)$ along with the knowledge that only even-$l$ operators appear in the $\phi \times \phi$ OPE.  It can similarly be verified that other crossings do not contain any additional information.

Note that by adding together Eqs.~(\ref{eq:crossingall}) and~(\ref{eq:crossingphiphis}) all of the odd-spin terms cancel and we recover the crossing symmetry constraint for the operators appearing in the $\Re[\phi] \times \Re[\phi]$ OPE.  Alternatively, we could subtract Eq.~(\ref{eq:crossingphiphis}) from Eq.~(\ref{eq:crossingall}) to obtain a crossing constraint which relates just the odd-spin operators appearing in $\phi \times \phi^{*}$ to the operators appearing in $\phi \times \phi$.  Also notice that the $(-1)^l$ factor in Eq.~(\ref{eq:crossingphiphis}) cancels against the $(-1)^l$ factor that occurs in the definition of the conformal blocks, so that in this equation the odd-spin terms are qualitatively similar to the even-spin terms.  This may be contrasted with Eq.~(\ref{eq:crossingall}), where odd-spin terms have the opposite sign relative to the even-spin terms.  For this reason, we have found that it is much easier to obtain a well-behaved linear program using the constraints of Eq.~(\ref{eq:crossingphiphis}) as compared to the constraints of Eq.~(\ref{eq:crossingall}).  Thus, in the present work we will mainly focus on the bounds that can be obtained using Eq.~(\ref{eq:crossingphiphis}), though in future studies it may be useful to incorporate the full set of constraints.

Let us also briefly mention another way to generalize the procedure outlined in the previous section.  Thus far, we have only used the knowledge that the unit operator appears in the crossing relation, but in many situations one might have additional information.  For example, if it is known that an operator $\tl \cO$ of dimension $\tl \De$ and spin $\tl l$ appears in the $\f \times \f$ OPE in addition to the unit operator, and we also know its three-point function coefficient $\l_{\tl \cO}$, then one can simply make the replacement $1 \rightarrow 1 - \l_{\tl \cO}^2 F_{\tl \De,\tl l}$ in the objective function of the linear program.  This modification can then lead to more stringent bounds.  It is particularly straightforward to implement in the case of the stress tensor $T^{a b}$ or a conserved global symmetry current $J^a$, since in these cases the dimensions are known and the $\l$'s are fixed by Ward identities, as we will review in Section~\ref{sec:bounds}.

\section{Superconformal Blocks}
\label{sec:scftblocks}

At this stage we could proceed to derive bounds on 3-point function coefficients in conformal field theories with global $U(1)$ symmetries.  However, because we would also like to derive similar bounds in $\cN=1$ superconformal theories, we will first consider more carefully the additional constraints imposed by supersymmetry.  In particular, three-point functions of primary operators in the same supersymmetry multiplet are related to each other by the superconformal algebra, and one can construct ``superconformal blocks" which sum up the contributions of all operators in a given superconformal multiplet.

We will focus on four-point functions involving a complex scalar $\f$ that is the lowest component of a chiral superfield $\Phi$ of dimension $d=\frac{3}{2} R_{\Phi}$.  In terms of the operators appearing in the $\phi \times \phi^*$ OPE, the superconformal block decomposition looks like
\ben
\label{eq:superconformalblockexpansion}
\<\f(x_1) \f^*(x_2) \f(x_3) \f^*(x_4)\> &=& \frac{1}{x_{12}^{2 d} x_{3 4}^{2 d}} \sum_{\mathcal{O} \in \Phi \times \Phi^\dag } |\lambda_{\mathcal{O}}|^2 (-1)^l \mathcal{G}_{\De,l}(u,v) .
\een
Here, we have adopted the notation $\cO\in \Phi\x\Phi^\dag$ to indicate that the sum is over {\it superconformal} primaries $\cO$ appearing in $\f\x\f^*$, and not simply primaries under the conformal subgroup.  By definition, superconformal primary operators $\cO$ are annihilated by the $S$ and $\bar{S}$ generators in the superconformal algebra, from which it follows that they are also annihilated by the $K$ generator.  However, a finite number of superconformal descendants of $\mathcal{O}$ are also killed by $K$, so one may decompose $\mathcal{G}_{\De,l}(u,v)$ into a finite sum of conformal blocks $g_{\De,l}(u,v)$.

Just as the explicit expression (\ref{eq:explicitconformalblocks}) for conformal blocks was crucial for the analysis of \cite{Rattazzi:2008pe,Rychkov:2009ij,Caracciolo:2009bx}, an explicit expression for superconformal blocks will be crucial for us.  We find that $\cN=1$ superconformal blocks in the $\f\x\f^*$ channel are given by
\ben
\label{eq:N=1superconformalblockintro}
\mathcal{G}_{\De,l}&=& g_{\De,l}-\frac{(\De+l)}{2(\De+l+1)}g_{\De+1,l+1}-\frac{(\De-l-2)}{8(\De-l-1)}g_{\De+1,l-1}\nonumber\\
&&\qquad\qquad+\frac{(\De+l)(\De-l-2)}{16(\De+l+1)(\De-l-1)}g_{\De+2,l} .
\een
To our knowledge, this expression has not yet appeared in the literature, though analogous results for $\cN=2$ and $\cN=4$ theories are known \cite{Dolan:2001tt}.  Eq.~(\ref{eq:N=1superconformalblockintro}) is the key ingredient we need to apply the technology of Section~\ref{sec:boundsfromcrossing} to superconformal theories.  In the following subsections, we will give two derivations --- one involving explicit analysis of superconformal two- and three-point functions, and another quicker but less illuminating argument leveraging known expressions from $\cN=2$ theories \cite{Dolan:2001tt}.  The discussion is somewhat technical, and readers interested solely in bounds on dimensions and OPE coefficients should feel free to skip to Section~\ref{sec:bounds}.

Our first derivation of Eq.~(\ref{eq:N=1superconformalblockintro}) proceeds as follows.  We start by understanding which superconformal primary operators $\cO^{a_1\dots a_l}$ can appear in the OPE $\f\x\f^*$.  We then determine which superconformal descendants of $\cO^{a_1\dots a_l}$ are conformal primaries, and further calculate the relationships between two- and three- point functions of these conformal primaries.  Since each conformal primary contributes a block $g_{\De',l'}$ to $\<\f\f^*\f\f^*\>$, we can piece together $\cG_{\De,l}$ from these contributions.  For completeness we also include a brief discussion of the $\f\x\f$ channel.  However, in this case only a single operator in each supersymmetry multiplet may contribute, so the superconformal blocks turn out to be the same as the conformal blocks  Eq.~(\ref{eq:explicitconformalblocks}).  Our conventions for the superconformal algebra and spinor notation are summarized in Appendix~\ref{app:conventions}.

\subsection{Superconformal Three-Point Functions}
\label{sec:supersymmetric3ptfns}

\subsubsection{$\phi \x \phi$ OPE}

Let us start by examining the $\phi \times \phi$ OPE, since the constraints from superconformal symmetry are particularly transparent in this case.  
This analysis is not needed later, but we include it for completeness and to establish some notation.  For some previous discussions of this OPE, see~\cite{Howe:1996rb,Dolan:2000uw}.
In this subsection we will follow the notation and conventions of~\cite{Osborn:1998qu}, where a superconformal primary $\cO^I$  ($I$ denotes Lorentz indices) is specified by spins $(j,\bar \jmath)$ and conformal weights $(q_\cO,\bar q_\cO)$, which are related to the dimension and $R$-charge via $q_\cO+\bar q_\cO=\De_{\cO}$ and $\frac 2 3(q_\cO-\bar q_\cO)=R_\cO$.  The unitarity bound for non-chiral superconformal primary operators then requires~\cite{Flato:1983te,Dobrev:1985qv,Minwalla:1997ka}
\ben
\label{eq:scunitaritybound}
\Delta_{{\mathcal O}^I} &\geq& |\frac32 R_{{\mathcal O}^I} - j+\bar{j}| + j + \bar{j} + 2.
\een

To begin, note that since $\bar{Q}\phi(x)=0$, only operators that are annihilated by $\bar Q$ may appear in $\phi \x \phi$.  A priori, there are four possibilities:
\begin{enumerate}
\item Chiral primaries.  Since these transform in $(j,0)$ representations of the Lorentz group $\SU(2)\x\SU(2)$, they can appear only if $j=0$.  We will denote the linear combination of chiral primaries appearing in $\f\x\f$ by $\f^2$.
\item Descendants of the form $\bar Q^{(\dot\a_1} \cO^{\dot\a_2\dots\dot\a_l)\a_1\dots \a_l}$, where $l$ is even and $\cO^I$ satisfies the shortening condition $\bar Q_{\dot\a}\cO^{\dot\a \dot \a_3\dots \dot \a_l \a_1\dots \a_l}=0$.  (Note that this implies $\bar Q^2\cO^I=0$, so that $\bar Q\cO^I$ is indeed killed by $\bar Q$.)  The superconformal algebra implies \cite{Osborn:1998qu} that such operators satisfy $\bar q_{\cO}=(l+1)/2$.  Then using $R_{\bar Q\cO^I}=2R_\Phi$ we find $\De_{\cO^I}=2d+l-\frac 1 2$, so that the dimensions of these operators are determined by their spins.  We will denote the linear combination of these descendants with spin $l$ as $\bar Q\cO_l^I$.\footnote{We are grateful to Alessandro Vichi for pointing out the possibility of these operators in the $\phi\x\phi$ OPE.}
\item Descendants of the form $\bar Q_{\dot\a} \cO^{\dot\a\dot\a_1\dots \dot\a_l\a_1\dots\a_l}$, where $\cO$ satisfies the shortening condition $\bar Q^{(\dot \a_1}\cO^{\dot\a_2\dots\dot\a_{l+2})\a_1\dots\a_l}=0$.  Such multiplets must satisfy $\bar q_{\cO}=-(l+1)/2$, which implies upon matching $R$-charges that $\De_\cO=2d-l-5/2$.  However, this violates the unitarity bound Eq.~(\ref{eq:scunitaritybound}), so such operators actually cannot appear.
\item Descendants of the form $\bar Q^2 Q^{2-n} \cO^I$, with $n = 0,1,2$.
\end{enumerate}
Thus, we expect the OPE to take the form
\ben
\label{eq:generalphiphiOPE}
\phi(x) \phi(0) &=& C(x,P) \f^2(0) + \sum_{l=2,4,\dots}C^l_I(x,P) \bar Q \cO_l^I(0)+\sum_{{\mathcal O}^I} \bar{Q}^2 C_I(x,P,Q) {\mathcal O}^I(0),
\een
where the latter sum runs over superconformal primaries with $R_{{\mathcal O}^I} = 2 R_{\Phi} - n$, and a priori $n=0,1,2$ depending on how many powers of $Q$ appear.

We can obtain additional constraints on the operators $\cO^I$ by acting on both sides of Eq.~(\ref{eq:generalphiphiOPE}) with an $S$ generator. Note that $S$ kills the left-hand side because $[S,P] \sim \bar{Q}$ and $\f$ is chiral and primary.  On the right-hand side, we can commute $S$ through all powers of $\bar{Q}$ and $P$, since  $\{S,\bar Q\}=0$ and $\bar Q^2[S,P] \sim\bar Q^3=0$.  However, if powers of $Q$ were present, there would be terms involving $\{S,Q\}$ which would not vanish when acting on ${\mathcal O}^I$.  Thus, we conclude that $C_I(x,P,Q) = C_I(x,P)$ and therefore $R_{{\mathcal O}^I} = 2 R_{\Phi} - 2$.  In this case the $I$ indices must correspond to even-spin operators due to the symmetry under exchanging $x \leftrightarrow -x$.  Finally, the unitarity bound Eq.~(\ref{eq:scunitaritybound}) implies
$
\Delta_{\bar{Q}^2 {\mathcal O}^I} \geq |3 R_{\Phi} - 3| + l + 3.
$
Note also that $\bar{Q}^2 \cO^I$ is primary under the conformal sub-algebra.

Instead of playing directly with the superconformal generators, an alternative approach that will prove useful later is to consider the general form of superconformal-covariant three-point functions.  Let us take a moment to recover the above results using this language.  

The $\Phi\x\Phi$ OPE contains a superconformal multiplet $\cO^{I}$ if and only if the three-point function $\<\Phi(z_{1+})\Phi(z_{2+})\cO^{I\dag}(z_3)\>$ is non-vanishing, where the $z$'s are superspace coordinates $(x,\th,\bar\th)$, and $z_+$ indicates dependence only on the chiral subspace $(x+i\th\s\bar\th,\th)$.  The general form of such a three-point function consistent with superconformal symmetry is
\ben
\label{eq:PhiPhiOthreeptfn}
\< \Phi(z_{1+}) \Phi(z_{2+}) {\mathcal O}^{I\dagger}(z_3) \> &=& \frac{t^I(X_3, \Theta_3,\bar{\Theta}_3)}{x_{\bar{3} 1}^{2d} x_{\bar{3} 2}^{2d}},
\een
where $x_{\bar{i}j} = x_{i-} + 2i\th_j \s \bar\th_i - x_{j+}$ denotes the supertranslation-invariant interval built out of anti-/chiral coordinates $x_{i\pm} = x_i \pm i\th_i\s\bar\th_i$, $X_3$ and $\Theta_3$ are given by
\ben
X_3^a &=& -\frac 1 2 \frac{x^b_{3\bar 1} x^c_{\bar 1 2} x^d_{2\bar 3}}{x_{\bar 1 3}^2x_{\bar 3 2}^2}\tr(\bar\s^a\s_b\bar\s_c\s_d),\\
\Theta_{3} &=& i\frac {x_{3\bar 1}^a} {x_{\bar 1 3}^2}\s_a\bar\th_{31}-i\frac {x_{3\bar 2}^a}{x_{\bar 2 3}^2}\s_a\bar\th_{3 2}, \qquad\bar\Theta_3=\Theta_3^\dagger,
\een
and $t^I$ has the homogeneity properties 
\ben
\label{eq:homogeneityproperty}
t^I(\lambda \bar{\lambda}X_3, \lambda \Theta_3, \bar{\lambda} \bar{\Theta}_3) &=& \lambda^{2 a} \bar{\lambda}^{2 \bar{a}} t^I(X_3, \Theta_3, \bar{\Theta}_3)
\een
with $a = \frac13 (2 q_{\cO} + \bar{q}_{\cO} - 4d)$ and $\bar{a} = \frac13(q_{\cO} + 2\bar{q}_{\cO} -2 d)$.

Since the covariant derivative $\bar{D}_1^{\dot{\a}}$ vanishes when acting on the left hand side of Eq.~(\ref{eq:PhiPhiOthreeptfn}), we obtain an additional constraint (using Eqs.~(6.1) and (6.2) in~\cite{Osborn:1998qu})
\ben
0 &=&\bar{D}_1^{\dot{\alpha}} t^I(X_3, \Theta_3, \bar{\Theta}_3)\nonumber\\
&=& - i \frac{(\mathrm{x}_{\bar{1} 3})^{\dot{\alpha} \alpha}}{x_{\bar{3} 1}^2} \left( \frac{\partial}{\partial \Theta_3^{\alpha}} - 2 i (\sigma^a \bar{\Theta}_3)_{\alpha} \frac{\partial}{\partial X_3^a}\right) t^I(X_3,\Theta_3,\bar{\Theta}_3) ,
\een
which implies that $t^I(X_3,\Theta_3,\bar{\Theta}_3) = t^I(\bar{X}_3,\bar{\Theta}_3)$, where $\bar{X}_3 \equiv X_3 + 2i \Theta_3 \sigma \bar{\Theta}_3$.  Finally, under $z_1 \leftrightarrow z_2$ we have $X_3 \leftrightarrow - \bar{X}_3$ and $\bar{\Theta}_3 \leftrightarrow - \bar{\Theta}_3$.  There are three possible solutions to these constraints,
\ben
t^I(\bar{X}_3,\bar{\Theta}_3) = \mathrm{const.},
\een
corresponding to $\cO^I$ being a chiral ``$\Phi^{2}$" operator with $R_{\cO} = 2 R_{\Phi}$,
\ben
t^I(\bar X_3,\bar \Theta_3) &\propto& \bar\Theta_3^{(\dot\a_1}\bar {X_3}^{\dot\a_2}_{\ \a_2}\dots\bar{X_3}^{\dot \a_l)}_{\ \a_l} \ \ =\ \  \bar\Theta_3^{(\dot\a_1} {X_3}^{\dot\a_2}_{\ \a_2}\dots{X_3}^{\dot \a_l)}_{\ \a_l},
\een
corresponding to the short operators $\cO_l^I$, and
\ben
t^I(\bar{X}_3,\bar{\Theta}_3) &\propto& \bar{\Theta}_3^2 \bar{X}_3^{\De_\cO-2d-l-1} \bar{X}_3^{a_1}\dots\bar{X}_3^{a_l}
\ \ =\ \ \bar{\Theta}_3^2 X_3^{\De_\cO-2d-l-1} X_3^{a_1}\dots X_3^{a_l},
\een 
corresponding to $\cO^I$ being a non-chiral operator with $R_{\cO} =  2 R_{\Phi}-2$.  Since the only irreducible Lorentz representations that can be built out of a single vector $\bar X_3^a$ (or $X_3^a$) are traceless symmetric tensors, $\cO^I=\cO^{a_1\dots a_l}$ must have definite integer spin $l=2j=2\bar{j}$, and invariance under $z_1 \leftrightarrow z_2$ further tells us that $l$ must be even.  The descendant operator $\bar{Q}^2 \cO^I$ then has the correct quantum numbers to appear in the $\phi\times\phi$ OPE, in precise agreement with the preceding argument.

Here we see that for each supermultiplet appearing in $\Phi \x \Phi$, there is exactly one conformal primary appearing in $\f \x \f$.  This is essentially because $\f^2$, $\bar Q \cO^I_l$, and $\bar Q^2 \cO^I$ are the only conformal primaries in their respective supermultiplets with the correct $R$-charge.  Consequently, the superconformal blocks for decomposing $\<\f\f^*\f\f^*\>$ in the $\f\x\f$ channel are the same as the conformal blocks.   Next we will turn to considering the $\f \x \f^*$ channel, where this will no longer be the case.

\subsubsection{$\f\x\f^*$ OPE}

We determine which operators can appear in the $\phi \x \phi^*$ OPE by examining three-point functions $\<\Phi \Phi^{\dagger} \cO^{I\dagger}\>$.  Once again, let $\cO^I$ be a superconformal primary with conformal weights $(q_\cO,\bar q_\cO)$ and spins $(j,\bar j)$.  Following~\cite{Osborn:1998qu}, we must have
\ben
\<\Phi(z_{1+})\Phi^\dag(z_{2-})\cO^{I\dagger}(z_3)\> &\propto& \frac{1}{x_{\bar 3 1}^{2d} x_{\bar 2 3}^{2d}}t^{I}(X_3,\Theta_3,\bar\Theta_3),
\een
where $t^I$ satisfies Eq.~(\ref{eq:homogeneityproperty}) with $a=\frac 1 3(2 q_\cO+ \bar q_\cO)-d$ and $\bar a=\frac 1 3(2 \bar q_\cO+ q_\cO)-d$.

Demanding the appropriate chirality properties imposes further constraints.  Just as in the $\f\x\f$ case, requiring $\bar D_1^{\dot{\a}} t^I=0$ means $t^I$ must be a function of $\bar X_3$ and $\bar \Theta_3$.  We must additionally require
\ben
0\ \ =\ \ D_{2}^\a t^I(\bar X_3,\bar\Theta_3) &=& i\frac{x_{2\bar 3}^{\dot\a \a}}{x_{\bar 2 3}^2}\pdr{}{\bar\Theta_{3}^{\dot\a}}t^I(\bar X_3,\bar\Theta_3),
\een
so that $t^I$ is actually a function of $\bar X_3$ alone.  Note that since the $R$-charge of $\bar X_3$ vanishes, the $R$-charge of the correlator $\<\Phi\Phi^\dag \cO^{I\dagger}\>$ must vanish as well, which means $\cO^{I}=\cO^{I\dag}$ should be a real operator with $q_\cO=\bar q_\cO$.  Since we again can only build Lorentz representations out of a single vector $\bar X_3^a$, the only possibilities are traceless symmetric tensors, so $\cO^I=\cO^{a_1\dots a_l}$ must have definite integer spin $l=2j=2\bar{j}$.

In summary, we have found that the only superconformal primaries appearing in the $\Phi\x\Phi^\dag$ OPE are traceless symmetric tensors $\cO^{a_1\dots a_l}$ with vanishing $R$-charge.  Superconformal symmetry determines the 3-point function to be
\ben
\label{eq:superconformal3ptfn}
\<\Phi(z_{1+})\Phi^\dag(z_{2-})\cO^{a_1\dots a_l}(z_3)\> &\propto& \frac 1 {x_{\bar 3 1}^{2d}x_{\bar 2 3}^{2d}}\bar X_3^{\De_\cO-2d-l}\bar X_3^{a_1}\dots \bar X_3^{a_l} - \mathrm{traces}.
\een
In this case, the unitarity bound Eq.~(\ref{eq:scunitaritybound}) requires $\De_\cO \geq l+2$.\footnote{With an exception, of course, for the unit operator which has $\De=l=0$.}  The operators which can enter the OPE of the lowest components $\f\x\f^*$ are then $R$-charge zero descendants of a real superconformal primary, $P^n(Q\bar Q)^m \cO^{a_1\dots a_l}$.  To understand how these operators contribute to the four-point function $\<\f\f^*\f\f^*\>$, we must now organize them into representations of the conformal sub-algebra.

\subsection{Decomposition of Superconformal Multiplets into Conformal Multiplets}

\label{sec:structureofN=1multiplets}

In this section, we will examine the structure of a multiplet built from a real superconformal primary $\cO^{a_1\dots a_l}$ of dimension $\De$.  The full superconformal multiplet can be decomposed into a direct sum of conformal multiplets, connected together by supersymmetry transformations.  Here we will show explicitly how this decomposition works for operators that appear in the $\phi\x\phi^*$ OPE --- namely operators of vanishing $R$-charge and definite spin.  As a result, we will see how superconformal symmetry relates the OPE coefficients of different conformal primaries, and consequently how $\cG_{\De,l}$ decomposes into a sum of $g_{\De,l}$'s.

Note that $\cO^{a_1\dots a_l}$ is symmetric and traceless in its indices.  Throughout this subsection, we will adopt the convention of implicitly symmetrizing and subtracting traces in $a_i$ for $i=1,\dots,l$.  This has the virtue of greatly simplifying notation, though one must be careful when manipulating expressions.
 
A convenient way to describe the descendants of a superconformal primary operator $\mathcal{O}^{a_1 \dots a_l}(0)$ is through superspace.  For example, defining the superfield $\mathcal{O}^{a_1 \dots a_l}(x,\th,\bar{\th}) = e^{x P + \th Q + \bar{\th} \bar{Q}} \mathcal{O}^{a_1 \dots a_l}(0)$, we have the component expansion
\ben
\label{eq:superfieldcomponents}
\cO^{a_1\dots a_l}(x,\th,\bar{\th}) &=& A^{a_1\dots a_l}(x)+\z_a B^{a a_1\dots a_l}(x)+\z^2 D^{a_1\dots a_l}(x)+\dots
\een
where $\z_a \equiv \th \sigma_a \bar{\th}$, and ``$\dots$" represents fields with non-zero $R$-charges.  The component fields $B^{a a_1 \dots a_l}$ and $D^{a_1 \dots a_l}$ are then related to $A^{a_1 \dots a_l}$ through the action of $Q$ and $\bar{Q}$ as
\ben
\label{eq:QQbaraction}
B^{a a_1 \dots a_l} &=& -\frac{1}{4} \Xi^a A^{a_1 \dots a_l}, \\
D^{a_1 \dots a_l} &=& -\frac{1}{64} \Xi_a B^{a a_1 \dots a_l} - \frac{1}{16} \ptl^2 A^{a_1 \dots a_l},
\een
where we have defined $\Xi^a \equiv \bar{\s}^{a \dot{\a} \a} [Q_{\a},\bar{Q}_{\dot{\a}}] $.

Both $A^{a_1 \dots a_l}$ and $D^{a_1 \dots a_l}$ are in the spin-$l$ representation of the Lorentz group, but $B^{a a_1 \dots a_l}$ can be further decomposed into irreducible representations. Recall that under $\SO(4)\cong\SU(2)\x\SU(2)$, the spin-$l$ representation of $\SO(4)$ transforms as $(j,j)$ with $j=l/2$.  Since $B^{a a_1 \dots a_l}$ has an additional vector index, it transforms as
\ben
\p{1/2,1/2}\otimes(j,j) &=& \phantom{\oplus}\p{j+1/2,j+1/2}\oplus\p{j-1/2,j-1/2}\nonumber\\
&&\oplus\p{j+1/2,j-1/2}\oplus \p{j-1/2,j+1/2}.
\een
The first two components on the right-hand side are a spin-$(l+1)$ representation $J^{aa_1\dots a_l}\equiv B^{(aa_1\dots a_l)}-\textrm{traces}$, and a spin-$(l-1)$ representation $N^{a_2\dots a_l}\equiv B_b{}^{ba_1\dots a_l}$.  The remaining two components comprise an operator $L^{aa_1\dots a_l}$ which is traceless and has vanishing total symmetrization.  $L$ can be further decomposed into irreducibles by projecting onto its ``anti/self-dual" parts, satisfying $L_{\pm}^{aa_1\dots a_l}=\pm i \frac l {l+1} \e^{aa_1}{}_{bc}L_{\pm}^{bca_2\dots a_l}$ (although this will not be important in our discussion).  Notice that since $L$ is not in a traceless symmetric representation, a primary operator built from it cannot appear in the OPE of $\f$ with $\f^*$.  Nonetheless, it will play a role in the identification of conformal primaries below.  Altogether, we may write
\ben
B^{aa_1\dots a_l} &=& J^{aa_1\dots a_l}+ \frac {l^2}{(l+1)^2}\eta^{aa_1}N^{a_2\dots a_l} +L^{aa_1\dots a_l},
\een
where as usual we are implicitly symmetrizing and subtracting traces in the $a_i$.  The coefficient of $N$ is such that the projection $N^{a_2\dots a_l}=B_b{}^{ba_1\dots a_l}$ works correctly.

Now let us consider the action of a special conformal generator $K_a$ on the components of $\mathcal{O}$.  We will be interested in determining which linear combinations of superconformal descendants are annihilated by $K_a$.  After some algebra, one can determine the action
\ben
K_a \left( \begin{array}{c} B^{b a_1 \dots a_l} \\ \e^{b a_1}{}_{c d} P^c A^{d a_2 \dots a_l} \end{array} \right) &=& \left( \begin{array}{c} 2l \\ 2(\De-1) \end{array}\right) \left(\e^{b a_1}{}_{a d} A^{d a_2 \dots a_l}\right),
\een
as well as
\ben
K_a \left( \begin{array}{c} D^{a_1 \dots a_l} \\ P^2 A^{a_1 \dots a_l} \\ P_b P^{a_1} A^{b a_2 \dots a_l} \\ \epsilon^{a_1 b}{}_{c d} P_b B^{c d a_2 \dots a_l} \end{array} \right) &=& \left( \begin{array}{cccc} \frac12 & 0 & 0 & \frac{l}{2} \\ 4(\Delta-1) & -4 l & 4 l & 0 \\ 0 & 2(\Delta-l-2) & 2(\Delta+l) & 0 \\ 2(l+1) & -2(l-1) & -2(l+1) & 2(\Delta-1) \\ \end{array} \right) \left( \begin{array}{c} P_a A^{a_1 \dots a_l} \\ P^{a_1}  A_a{}^{a_2 \dots a_l} \\ \de^{a_1}_{a} P_b A^{b a_2 \dots a_l} \\ \epsilon^{a_1}{}_{acd}B^{cd a_2 \dots a_l} \end{array} \right), \nonumber\\
\een
from which we find that the linear combinations
\ben
\label{eq:primarycombinations}
B_{\mathrm{prim}}^{a a_1 \dots a_l} &\equiv&  B^{a a_1 \dots a_l} - \frac{l}{\De-1} \e^{a a_1}{}_{cd}P^c A^{d a_2\dots a_l}\\
D_{\mathrm{prim}}^{a_1 \dots a_l} &\equiv& D^{a_1 \dots a_l} + \frac{l(l+1)-(\Delta-1)}{8 (\Delta-1)^2} P^2 A^{a_1 \dots a_l} - \frac{l^2}{4 (\Delta-1)^2} P_b P^{a_1} A^{b a_2 \dots a_l}\nonumber\\
&& - \frac{l}{4(\Delta-1)}\epsilon^{a_1 b}{}_{c d} P_b B^{c d a_2 \dots a_l}
\een
are primary operators under the conformal subgroup.  Note that only the $L$ component of $B$ is shifted in the above expression for $B_{\mathrm{prim}}$, so that $J$ and $N$ are already primary.

An important fact is that when the unitarity bound $\De\geq l+2$ is saturated, our superconformal multiplet is ``shortened," and the descendants $N,L_\prim$, and $D_\prim$ actually vanish.  For example, the supercurrent $\cJ^a(z)$ with $\De=3$ and $l=1$ contains only the $R$-symmetry current $J^a_R(x)$ and stress tensor $T^{ab}(x)$ as conformal primary components with vanishing $R$-charge.  This will be reflected in explicit calculations below.

\subsection{Conformal Primary Three-Point Functions}
Next we would like to see how the three point functions $\<\phi \phi^{*} J\>$, $\<\f \f^{*} N\>$, and $\<\f \f^{*} D_{\mathrm{prim}}\>$ are related to $\<\f \f^{*} A\>$.  We will also verify that $\<\f \f^{*} L_{\mathrm{prim}}\> = 0$, as expected because $L_{\mathrm{prim}}$ is not in an integer-spin (traceless symmetric) representation of the Lorentz group.

Let us set $\th_1=\th_2=\bar\th_1=\bar\th_2=0$, and $\th_3=\th, \bar\th_3=\bar\th$ in the correlator Eq.~(\ref{eq:superconformal3ptfn}) to get the 3-point function $\<\f(x_1) \f^{*}(x_2)\cO^{a_1 \dots a_l}(x_3,\th,\bar\th)\>$.  Next, expanding in $\th,\bar\th$ and comparing with our component expansion Eq.~(\ref{eq:superfieldcomponents}), we find
\ben
\<\f\f^{*} A^{a_1\dots a_l}\> &=& \frac{x_{12}^{\De-2q-l}}{x_{13}^{\De-l}x_{23}^{\De-l}} Z^{a_1}\dots Z^{a_l}\\
\<\f\f^{*}  J^{aa_1\dots a_l}\> &=& i(\De+l)\frac{x_{12}^{\De-2q-l}}{x_{13}^{\De-l}x_{23}^{\De-l}}Z^a Z^{a_1}\dots Z^{a_l}\\
 \<\f\f^{*} N^{a_2 \dots a_l} \>&=& i \frac{(\De-l-2)(l+1)}{2l} \frac{x_{12}^{\De-2q-l}}{x_{13}^{\De-l}x_{23}^{\De-l}} Z^2 Z^{a_2} \dots Z^{a_l} \\
\<\f\f^{*}  L^{aa_1\dots a_l}\> &=& 2l \frac{x_{12}^{\De-2q-l}}{x_{13}^{\De-l}x_{23}^{\De-l}} Y^{aa_1}Z^{a_2}\dots Z^{a_l}\\
\<\f\f^{*}  D^{a_1\dots a_l}\> &=& \frac{x_{12}^{\De-2q-l}}{x_{13}^{\De-l}x_{23}^{\De-l}}\left(\frac l 2 Z^2 \frac{x_{21}^{a_1}}{x_{21}^2} Z^{a_2}\dots Z^{a_l}+\frac{l(l-1)}{2}\eta_{bc}Y^{a_1b}Y^{a_2c} Z^{a_3}\dots Z^{a_l}\right.\\
&&\left.\qquad-\p{\frac{(\De+l)}{2}\frac{x_{32}\.x_{13}}{x_{13}^2 x_{32}^2}+\frac{(\De+l)(\De-l-2)}{8}Z^2} Z^{a_1}\dots Z^{a_l}\right),\nonumber
\een
where
\ben
Z^a \equiv \frac{x_{31}^a}{x_{31}^2}-\frac{x_{32}^a}{x_{32}^2},\qquad
Y^{ab} \equiv \frac{1}{x_{32}^2 x_{31}^2}\e^{ab}{}_{cd}x_{31}^cx_{32}^d,
\een
and we are implicitly projecting the right-hand side of each expression onto the appropriate Lorentz representation (symmetrizing and subtracting traces as necessary).  Using $Z^2=x_{12}^2/(x_{31}^2x_{32}^2)$, we see that the correlators $\<\f\f^*A\>,\<\f\f^*J\>$ and $\<\f\f^*N\>$ take the expected form for a 3-point function of conformal primary operators.  Further, taking the appropriate derivatives of the above expressions and constructing the linear combinations corresponding to $L_\prim$ and $D_\prim$, we obtain
\ben
\<\f\f^{*}  L_{\mathrm{prim}} ^{aa_1\dots a_l}\> &=& 0
\een
as expected, and
\ben
\<\f\f^{*} D_{\mathrm{prim}}^{a_1\dots a_l}\> &=& -\frac{\De(\De+l)(\De-l-2)}{8(\De-1)} \frac{x_{12}^{\De-2q-l}}{x_{13}^{\De-l}x_{23}^{\De-l}} Z^2 Z^{a_1}\dots Z^{a_l} .
\een
Notice that three-point functions involving $N$ and $D_{\mathrm{prim}}$ vanish when $\De = l+2$, which is precisely what we expect for short multiplets that saturate the unitarity bound. 

\subsection{Conformal Primary Norms}
Finally we must determine the normalization of the two-point functions $\<J J\>$, $\<N N\>$, and $\<D_{\mathrm{prim}} D_{\mathrm{prim}}\>$.  One could do this either by expanding out the superconformally covariant expression for the two-point function of $\mathcal{O}$ derived in~\cite{Osborn:1998qu} into its various components, or by using the explicit expressions for $J,N$, and $D_{\mathrm{prim}}$ in terms of $Q,\bar{Q}$, and $P$ acting on $A$, and using the superconformal algebra to compute their norms in radial quantization.  We here adopt the latter approach.  We refer the reader to \cite{Minwalla:1997ka} for many examples of this type of computation.  

To begin, we assume that the superconformal primary operator $A$ is canonically normalized
\ben
\<A^{b_1\dots b_l} | A^{a_1\dots a_l}\> &=& \textrm{symmetrize}(\eta^{a_1b_1}\dots \eta^{a_lb_l})-\textrm{traces}\nonumber\\
&=& \frac 1 {l!}\sum_{\pi\in S_l}\eta^{a_1b_{\pi(1)}}\dots \eta^{a_l b_{\pi(l)}}-\textrm{traces}\nonumber\\
&\equiv& \mathcal{I}_l^{a_1\dots a_l; b_1\dots b_l},
\een
where we've defined $\mathcal{I}_l^{a_1\dots a_l; b_1\dots b_l}$ for future convenience, and $| A^{a_1 \dots a_l} \> = A^{a_1 \dots a_l}(0) |0\>$ is the state created by the operator $A^{a_1 \dots a_l}(0)$ in radial quantization.

Next we would like to determine the normalization of $B_{\mathrm{prim}}^{a a_1 \dots a_l}$.   Starting from Eqs.~(\ref{eq:QQbaraction}) and~(\ref{eq:primarycombinations}) and working through the algebra, we find that 
\ben
\<B_{\mathrm{prim}}^{b b_1 \dots b_l} | B_{\mathrm{prim} }^{a a_1 \dots a_l} \> &=& 2 \left( \left(\Delta (\Delta+1)-l^2 - \frac{l(l+1)}{\De-1} \right) \eta^{b a} \eta_{c_1}^{a_1}+ l \left( 2\Delta+2l+1+ \frac{l+1}{\De-1}  \right) \eta^{b a_1} \eta^a_{c_1} \right.\nonumber \\
 &&\qquad\qquad \left.  - l \left(2\Delta-2l+1-\frac{l-1}{\De-1}\right) \eta^b_{c_1} \eta^{a a_1}  \right)  \< A^{b_1 \dots b_l} | A^{c_1 a_2 \dots a_l} \> ,
\een
from which we can extract the component normalizations
\ben
\<J^{b b_1 \dots b_l} | J^{a a_1 \dots a_l}\> 
&=& 2(\Delta+l)(\Delta+l+1) \mathcal{I}_{l+1}^{aa_1\dots a_l;bb_1\dots b_l} ,
\een
as well as
\ben
\<N^{b_2 \dots b_l} | N^{a_2 \dots a_l}\> 
&=&  \frac{2(l+1)^2}{l^2}(\Delta-l-2)(\Delta-l-1)\mathcal{I}_{l-1}^{a_2\dots a_l;b_2\dots b_l} ,
\een
where we have used the relation 
$ \eta_{ab}\mathcal{I}_{l}^{aa_2\dots a_l;bb_2\dots b_l} = \frac{(l+1)^2}{l^2}\mathcal{I}_{l-1}^{a_2\dots a_l;b_2\dots b_l} $.
Although we will not need it, for completeness we also have
\ben
\<L_\mathrm{prim}^{b b_1 \dots b_l} | L_\mathrm{prim}^{a a_1 \dots a_l}\> 
&=&\frac{8l^2\De(\De+l)(\De-l-2)}{(l+1)^2(\De-1)} \eta^{a b}  \mathcal{I}_{l}^{a_1\dots a_l;b_1\dots b_l} ,
\een
where we are implicitly subtracting traces and the full symmetrization (in either the $a,a_i$ or $b,b_i$ indices) from the right hand side --- that is, projecting onto the Lorentz representation corresponding to $L$.

Finally we must determine the normalization of $D_{\mathrm{prim}}^{a_1 \dots a_l}$.  In order to simplify the calculation, 
it will be helpful to write everything in terms of primary fields, 
\ben
D^{a_1 \dots a_l}_\mathrm{prim} &=& -\frac{1}{64} \Xi_a B^{a a_1 \dots a_l}_{\mathrm{prim}} - \frac{l(l+1) +(\De-1)(\De+1)}{16(\De-1)^2} P^2 A^{a _1 \dots a_l} + \frac{l^2}{8(\De-1)^2}P^{a_1} P_b A^{b a_2 \dots a_l} \nonumber\\
&& -\frac{3l}{16(\De-1)} \e^{a_1b}{}_{cd}P_b B^{cda_2\dots a_l}_\mathrm{prim} \een
so that 
\ben
 64^2 \<D_{\mathrm{prim}}^{b_1 \dots b_l} | D_{\mathrm{prim}}^{a_1 \dots a_l} \>&=& 
 \<B_{\mathrm{prim}}^{b b_1 \dots b_l} | (\Xi_b)^{\dagger} \Xi_a | B_{\mathrm{prim}}^{a a_1 \dots a_l} \>
 - \frac{8 l^2}{(\De-1)^2} \<B_{\mathrm{prim}}^{b b_1 \dots b_l} | (\Xi_b)^{\dagger} P^{a_1} P_c |A^{c a_2 \dots a_l} \> \nonumber\\
&& + 4 \frac{l(l+1) +(\De-1)(\De+1)}{(\De-1)^2}  \<B_{\mathrm{prim}}^{b b_1 \dots b_l} | (\Xi_b)^{\dagger} P^2 | A^{a _1 \dots a_l}  \>  \nonumber\\
&& + \frac{12 l}{(\De-1)}  \e^{a_1e}{}_{cd}\<B_{\mathrm{prim}}^{b b_1 \dots b_l} | (\Xi_b)^{\dagger}  P_e | B^{cda_2\dots a_l}_\mathrm{prim}\>,
\een
where we have used that all terms of the form $\< (\dots) K | D_{\mathrm{prim}}\>$ vanish.  Evaluating each of these terms using the superconformal algebra and putting everything together, we obtain the final result
\ben
\<D_{\mathrm{prim}}^{b_1 \dots b_l } | D_{\mathrm{prim}}^{a_1 \dots a_l}\> 
&=&\frac{\De^2 (\De-l-2)(\De-l-1)(\De+l)(\De+l+1)}{4(\De-1)^2} \mathcal{I}_{l}^{a_1\dots a_l;b_1\dots b_l} .
\een

\subsection{$\mathcal{N}=1$ Superconformal Blocks}
\label{sec:N=1superconformalblocksresult}
To summarize the results in the previous subsections, we have found the three-point function coefficients
\ben
\l_{\f\f^* A} &=& 1\nonumber\\
\l_{\f\f^* J} &=& i(\De+l)\nonumber\\
\l_{\f\f^* N} &=& i\frac{(\De-l-2)(l+1)}{2l}\nonumber\\
\l_{\f\f^* D} &=& -\frac{\De(\De+l)(\De-l-2)}{8(\De-1)}
\een
and the norms
\ben
\<A | A\> &\sim& 1\nonumber\\
\<J | J\> &\sim& 2(\De+l)(\De+l+1)\nonumber\\
\<N | N\> &\sim& \frac{2(l+1)^2(\De-l-2)(\De-l-1)}{l^2}\nonumber\\
\<D | D\> &\sim& \frac{\De^2 (\De-l-2)(\De-l-1)(\De+l)(\De+l+1)}{4(\De-1)^2} ,
\een
where ``$\sim$" means multiplied by the appropriate canonically normalized tensor.  Combining these results, we find the dimension $\De$, spin $l$ superconformal block given in Eq.~(\ref{eq:N=1superconformalblockintro}), which we reproduce here for the reader's convenience,
\ben
\label{eq:N=1superconformalblock}
\mathcal{G}_{\De,l}&=& g_{\De,l}-\frac{(\De+l)}{2(\De+l+1)}g_{\De+1,l+1}-\frac{(\De-l-2)}{8(\De-l-1)}g_{\De+1,l-1}\nonumber\\
&&\qquad\qquad+\frac{(\De+l)(\De-l-2)}{16(\De+l+1)(\De-l-1)}g_{\De+2,l} .
\een

A few comments are in order.  First, $l=0$ is special, since in this case the $N$ component does not exist.  However, one can consistently take $g_{\De,-1}=0$, and then the above equation correctly accounts for this situation.  Second, in the case of superconformal primary operators that saturate the unitarity bound, $\De=l+2$, the third and fourth terms vanish, which is precisely what  we expect due to the fact that the $N$ and $D_\prim$ components are not present in short multiplets.  Finally, in the case of the unit operator, with $\De=l=0$, the second and fourth terms vanish due to the coefficient going to zero, and the third term vanishes because the conformal block goes to zero.  Thus, we simply obtain that $\mathcal{G}_{0,0} = g_{0,0}=1$.

Let us also note that Eq.~(\ref{eq:N=1superconformalblock}) determines the superconformal blocks for four-point functions of all component fields in $\Phi(z_+)$, not just the lowest component $\f(x)$.  The reason is that there are unique superconformally-invariant extensions of the conformally-invariant cross-ratios $u,v$ with the correct chirality properties to appear in a four-point function $\<\Phi(z_{1+})\Phi^\dag(z_{2-})\Phi(z_{3+})\Phi^\dag(z_{4-})\>$.  They are given by \cite{Osborn:1998qu}
\ben
\tl u = \frac{x_{\bar 2 1}^2 x_{\bar 4 3}^2}{x_{\bar 2 3}^2 x_{\bar 4 1}^2},
\qquad
\tl v = \frac 1 2 \tr(x_{\bar 2 1}x_{\bar 4 1}^{-1}x_{\bar 4 3}x_{\bar 2 3}^{-1}),
\een
where the $x$'s in the trace should be thought of as bispinors, $(x)^{\dot\a\a}=x^a\bar\s_a^{\dot\a\a}$ and $(x^{-1})_{\a\dot\a}=-x_a \s^a_{\a\dot\a}/x^2$.  Since $\tl u$ and $\tl v$ become $u$ and $v$ when we set $\th_i=\bar\th_i= 0$, we must have
\ben
\label{eq:superspaceconformalblocks}
\<\Phi(z_{1+})\Phi^\dag(z_{2-})\Phi(z_{3+})\Phi^\dag(z_{4-})\> &=& \frac{1}{x_{1\bar 2}^{2d}x_{3\bar 4}^{2d}}\sum_{\cO\in\Phi\x\Phi^\dag}\l_\cO^2 \cG_{\De,l}(\tl u,\tl v),
\een
where $\cG_{\De,l}$ is given by Eq.~(\ref{eq:N=1superconformalblock}) above.  One can now perform $\th,\bar\th$ expansions on both sides to derive the superconformal blocks for specific component fields.

Finally, let us mention that it may be possible to derive the superconformal blocks by mimicking the derivation of $g_{\De,l}$ in \cite{Dolan:2003hv}.  One would start with the expansion Eq.~(\ref{eq:superspaceconformalblocks}) and apply the quadratic casimir of the superconformal group acting on $\Phi(z_{1+})$ and $\Phi(z_{2-})$ to obtain a differential equation for $\cG_{\De,l}$, which could then be solved.

\subsection{Deriving $\cN=1$ Blocks From $\mathcal{N}=2$ Blocks}

In~\cite{Dolan:2001tt}, Dolan and Osborn computed superconformal blocks for four-point functions of a particular kind of BPS operator in $\cN=2$ theories, using Ward identities special to higher supersymmetry.  At the very least, we should be able to decompose their expression into $\cN=1$ superconformal blocks $\cG_{\De,l}$.  However, requiring that this is possible gives a strong consistency condition on $\cG_{\De,l}$ --- so strong in fact that it determines $\cG_{\De,l}$ completely!  In this subsection, we will use this fact to give an alternate derivation of Eq.~(\ref{eq:N=1superconformalblock}) that requires far less computation than in Sections~\ref{sec:supersymmetric3ptfns}-\ref{sec:N=1superconformalblocksresult}, though it leverages important results from \cite{Dolan:2001tt}.

The operator $\vf^{ij}$ considered in~\cite{Dolan:2001tt} is a triplet under $\SU(2)_R$, neutral under $U(1)_R$, and has scaling dimension 2 (here $i,j=1,2$ are $\SU(2)_R$ indices).  It satisfies the BPS conditions $Q^{(i}_\a\vf^{jk)}=\e^{l(i}\bar Q_{\dot\a l}\vf^{jk)}=0$, which imply that under the $\cN=1$ sub-algebra generated by $Q^1_\a$ and $\bar Q_{\dot\a 1}$, the operators $\vf^{11}, \vf^{21}$, and $\vf^{22}$ are anti-chiral, linear, and chiral respectively.  The important fact for us is that $\vf^{22}\equiv \f$ is chiral, so $\<\f\f^*\f\f^*\>$ can be decomposed into a sum of $\cG_{\De,l}$'s.  Note that the form of $\cG_{\De,l}$ is independent of the dimension of $\f$.  In particular, it is irrelevant for our purposes that $\f$ is restricted to have dimension $2$.

Any $\cN=2$ multiplet that can appear in the OPE $\vf^{ij}\x \vf^{kl}$ must be built from a primary of dimension $\De$ and definite integer spin $l$.  We will denote such a multiplet by $(\De)_l^{\cN=2}$.  The ``extra" supersymmetry generators $Q^2, \bar Q_2$ connect different $\cN=1$ multiplets within $(\De)_l^{\cN=2}$ exactly analogously to the way $Q$ and $\bar Q$ connect different conformal multiplets within $(\De)_l^{\cN=1}$, as discussed in Section~\ref{sec:structureofN=1multiplets}.  Thus, we have the decompositions
\ben
(\De)_l^{\cN=2} &=& (\De)_l^{\cN=1}\oplus(\De+1)_{l\pm 1}^{\cN=1}\oplus (\De+2)_l^{\cN=1}\\
(\De)_l^{\cN=1} &=& (\De)_l^{\cN=0}\oplus(\De+1)_{l\pm 1}^{\cN=0}\oplus (\De+2)_l^{\cN=0},
\een
where we have ignored multiplets which cannot appear in the OPE of two scalars.  We can then write the ansatze
\ben
\label{eq:ansatzeforblocks1}
\cG^{\cN=2}_{\De,l}&=&\cG_{\De,l}+N(\De,l)\cG_{\De+1,l-1}+J(\De,l)\cG_{\De+1,l+1}+D(\De,l)\cG_{\De+2,l}
\\
\label{eq:ansatzeforblocks2}
\cG_{\De,l} &=& g_{\De,l}+n(\De,l)g_{\De+1,l-1}+j(\De,l)g_{\De+1,l+1}+d(\De,l)g_{\De+2,l},
\een
where $N,J,D,n,j,d$ are functions we would like to determine.  Note that $j,n$, and $d$ must be rational functions of $\De$ and $l$.  This is clear without any computation, simply from the viability of our first method for determining $\cG_{\De,l}$ (Sections~\ref{sec:supersymmetric3ptfns}-\ref{sec:N=1superconformalblocksresult}).

Using formulae from~\cite{Dolan:2001tt}, we find that the $\cN=2$ superconformal block contributing to $\<\f\f^*\f\f^*\>$ is given in terms of conformal blocks by
\ben
\label{eq:N=2superconformalblock}
\cG^{\cN=2}_{\De,l}
&=& g_{\De,l}-g_{\De+1,l+1}-\frac 1 4 g_{\De+1,l-1}+\frac 1 4 g_{\De+2,l}\nonumber\\
&&+\frac{(\De+l+2)^2}{4(\De+l+1)(\De+l+3)}g_{\De+2,l+2}-\frac{(\De+l+2)^2}{16(\De+l+1)(\De+l+3)}g_{\De+3,l+1}\nonumber\\
&&+\frac{(\De-l)^2}{64(\De-l-1)(\De-l+1)}g_{\De+2,l-2}-\frac{(\De-l)^2}{64(\De-l-1)(\De-l+1)}g_{\De+3,l-1}\nonumber\\
&&+\frac{(\De+l+2)^2(\De-l)^2}{256(\De+l+1)(\De+l+3)(\De-l-1)(\De-l+1)}g_{\De+4,l} .
\een

Upon comparison with Eqs.~(\ref{eq:ansatzeforblocks1}) and~(\ref{eq:ansatzeforblocks2}), each coefficient in the above expression implies an equation relating $N,J,D,n,j,$ and $d$.  We will solve these equations by first determining $j$ and $n$, and finally computing $d$ in terms of them.  To begin, the $g_{\De+1,l+1}$ and $g_{\De+2,l+2}$ terms in Eq.~(\ref{eq:N=2superconformalblock}) imply
\ben
\label{eq:eqnsforJj}
-1 = J(\De,l)+j(\De,l),\quad\textrm{and}\quad \frac{(\De+l+2)^2}{4(\De+l+1)(\De+l+3)}=J(\De,l)j(\De+1,l+1) .
\een
With some foresight, but without loss of generality, let us make the substitution
\ben
j(\De,l) &=& -\frac{(\De+l)}{2(\De+l+1)}(1+\a(\De+l,\De-l)),
\een
where $\a(x,\bar x)$ is a rational function we must determine. Then Eqs.~(\ref{eq:eqnsforJj}) imply the equation
\ben
\a(x,\bar x) &=& \frac{x+2}{x}\frac{\a(x+2,\bar x)}{1+\a(x+2,\bar x)},
\een
and it's not difficult to show that any rational solution $\a(x,\bar x)$ must vanish identically.  Consequently, we obtain
\ben
j(\De,l) = -\frac{(\De+l)}{2(\De+l+1)},\qquad J(\De,l)=-\frac{(\De+l+2)}{2(\De+l+1)} .
\een
A similar analysis using the $g_{\De+1,l-1}$ and $g_{\De+2,l-2}$ terms in Eq.~(\ref{eq:N=2superconformalblock}) gives
\ben
n(\De,l) = -\frac{(\De-l-2)}{8(\De-l-1)},\qquad N(\De,l) = -\frac{(\De-l)}{8(\De-l-1)}.
\een

Finally, let us solve for $d(\De,l)$.  The $g_{\De+4,l}$ term in Eq.~(\ref{eq:N=2superconformalblock}) determines $D(\De,l)$ in terms of $d(\De+2,l)$.  Plugging this in, along with our solutions for $N,J,n$, and $j$, the remaining terms in Eq.~(\ref{eq:N=2superconformalblock}) imply equations with the following structure
\ben
g_{\De+2,l}: && d(\De,l) \sim d(\De+2,l)\\
g_{\De+3,l+1}: && d(\De+1,l+1) \sim d(\De+2,l)\label{eq:eqntosubstitute}\\
g_{\De+3,l-1}: && d(\De+1,l-1) \sim d(\De+2,l),
\een
where ``$\sim$" means ``is algebraically related to."  Making the substitutions $\De\to\De-1$ and $l\to l-1$ in Eq.~(\ref{eq:eqntosubstitute}), we are left with three algebraic equations relating three ``variables" $d(\De,l), d(\De+2,l),$ and $d(\De+1,l-1)$.  Solving them gives
\ben
d(\De,l)=\frac{(\De+l)(\De-l-2)}{16(\De+l+1)(\De-l-1)},\qquad D(\De,l)=\frac{(\De+l+2)(\De-l)}{16(\De+l+1)(\De-l-1)}.
\een

To summarize, we have re-derived Eq.~(\ref{eq:N=1superconformalblock}),\footnote{It's possible that similar arguments suffice to determine $\cN=2$ conformal blocks from $\cN=4$ conformal blocks.  If this is the case, it's fascinating that a maximally supersymmetric result, which can be derived using special properties of $\cN=4$ BPS multiplets, completely determines the corresponding results for lower supersymmetry. } and also obtained the decomposition of $\cN=2$ conformal blocks into $\cN=1$ conformal blocks
\ben
\cG^{\cN=2}_{\De,l} &=& \cG_{\De,l}-\frac{(\De+l+2)}{2(\De+l+1)}\cG_{\De+1,l+1}-\frac{(\De-l)}{8(\De-l-1)}\cG_{\De+1,l-1}\nonumber\\
&&\qquad\qquad+\frac{(\De+l+2)(\De-l)}{16(\De+l+1)(\De-l-1)}\cG_{\De+2,l}.
\een

\section{Bounds}
\label{sec:bounds}
Now we finally turn to using the results obtained in Sections~\ref{sec:cft} and~\ref{sec:scftblocks} to obtain bounds on CFT and SCFT data.  We will start by considering bounds on the OPE coefficient of the lowest-dimension scalar appearing in the $\Phi \times \Phi^{\dagger}$ OPE (which we call ``$\Phi^{\dagger} \Phi$"), where $\Phi$ is a chiral multiplet in an $\cN=1$ superconformal theory.  When the dimension of $\Phi$ is somewhat close to $1$, we find that these OPE coefficient bounds are sufficiently strong to yield an upper bound on the dimension of $\Phi^{\dagger}\Phi$.  This is a completely general result about the dimensions of non-chiral operators in strongly-coupled $\cN=1$ superconformal field theories.  We will also present a bound on the OPE coefficient of an arbitrary scalar operator that can appear in this OPE, independent of any assumptions about the spectrum.

Then we turn to bounding the OPE coefficients of flavor currents.  In general CFTs these are spin-$1$ operators $J^{aI}$ of dimension $3$, and in $\cN=1$ theories the $J^{aI}$ are embedded into real scalar operators $J^I$ of dimension 2.  We will review how Ward identities fix these OPE coefficients in terms of the coefficient of $\<J^I J^J\> \propto \tau^{I J}$ and the charges of $\f$, allowing us to bound the quantity $\tau_{I J} T^I T^J$, where $\tau_{I J} = (\tau^{I J})^{-1}$ and $T^I$ are the generators of the flavor symmetry in the $\f$ representation.  Roughly speaking, $\tau^{IJ}$ measures the number of degrees of freedom charged under the global symmetries,\footnote{For example, if the flavor symmetry is weakly gauged with coupling $g$, then $\tau^{IJ}$ is proportional to the contribution of the CFT to the beta function of $g$.} and our bound says that the effective number of degrees of freedom that are charged cannot be much smaller than $1$.  We present this bound in both non-supersymmetric and supersymmetric CFTs.

Finally we consider the OPE coefficient of the stress tensor, which is similarly fixed by Ward identities in terms of the dimension $d$ of $\f$ and the central charge $c$.  This will allow us to derive a lower bound on the value of the central charge in both non-supersymmetric and supersymmetric CFTs.  In the former case, the stress tensor is a spin-$2$ operator of dimension 4, and the bound will assume only the existence of a real scalar primary operator of dimension $d$.  In the latter case, the stress-tensor is the $\th\s^a\bar\th$ component of a spin-$1$ $U(1)_R$ current multiplet of dimension $3$, and the bound will assume only the existence of a chiral primary scalar of dimension $d$.  

\subsection{Dimension of $\Phi^\dag\Phi$ and Scalar OPE Coefficients}
\label{sec:OPEcoeffbounds}

We start from the crossing relation Eq.~(\ref{eq:crossingphiphis}), which involves only the $\f\x\f^*$ channel of the four-point function $\<\f\f^*\f\f^*\>$.  Superconformal symmetry additionally allows us to group terms into superconformal blocks, so that we may write
\ben\label{eq:crossingphiphissuper}
\sum_{\cO \in \Phi \times \Phi^\dag} |\l_{\cO}|^2 (-1)^l \cG_{\De,l}(u,v) &=& \left(\frac{u}{v}\right)^{d} \sum_{\cO \in \Phi \times \Phi^\dag} |\l_{\cO}|^2 (-1)^l \cG_{\De,l}(v,u),
\een
where as before we have written $\Phi\x\Phi^\dag$ instead of $\f\x\f^*$ to indicate that the sum is over {\it superconformal} primaries in the OPE of $\f$ with $\f^*$, as opposed to simply primaries under the conformal subgroup.\footnote{Note that the methods in this section apply equally well to a general CFT with a global $U(1)$ symmetry, in which case the bounds are strictly weaker.}

From here, the procedure is exactly as described in Section~\ref{sec:boundsfromcrossing}.  Suppose the operator $\cO_0$ with dimension $\De_0$ is the lowest-dimension scalar appearing in $\Phi\x\Phi^\dag$.  Isolating the contributions of $\cO_0$ and the unit operator, we have
\ben
\label{eq:crossingrewrite}
|\l_{\cO_0}|^2\cF_{\De_0,l_0} &=& 1 - \sum_{\cO\neq \cO_0}|\l_{\cO}|^2 \cF_{\De,l} ,
\een
where $\cF_{\De,l}$ is given by Eq.~(\ref{eq:FdeltaL}) with $g_{\De,l}\to (-1)^l\cG_{\De,l}$.  
Finally, to obtain the best possible bound $|\l_{\cO_0}|^2\leq \a(1)$, we must minimize $\a(1)$ over all $\a\subset\cV^*$ satisfying the constraints
\begin{itemize}
\item $\a(\cF_{\De,0})\geq 0$ for all $\De\geq \De_0$,
\item $\a(\cF_{\De,l})\geq 0$ for all $\De\geq l+2$ and $l\geq 1$ (not necessarily even),
\item $\a(\cF_{\De_0,0})=1$.
\end{itemize}
Then if the resulting bound tells us that $|\l_{\cO_0}|^2 \leq \alpha(1) < 0$, there is a contradiction with unitarity, and we learn that $\cO_0$ cannot have dimension $\De_0$.   

Let us highlight some assumptions implicit in this procedure.  Firstly, we are assuming that $\Phi$ is uncharged under any global flavor symmetries (that is, non-$R$ symmetries), since otherwise there would be a symmetry current $J$ of dimension $2$ in the OPE $\Phi\x\Phi^\dag$, which would necessarily be the lowest-dimension scalar by the SUSY unitarity bound.  Alternatively, if $\Phi$ has flavor charges, and we wish to bound the lowest-dimension scalar in $\Phi\x\Phi^\dag$ that is {\it not} a flavor current, then we must incorporate flavor current blocks $\cF_{2,0}$ into the objective function of our linear program $1\to 1-|\l_J|^2\cF_{2,0}$, as discussed in Section~\ref{sec:limitationsofrattazzi}.

Secondly, note that we are only using part of the full crossing relation Eq.~(\ref{eq:crossingphiphi}), and it is possible that one could obtain stronger bounds by incorporating the additional relations Eq.~(\ref{eq:crossingall}) (whose terms also can be grouped into superconformal blocks if desired).  So far, we have not had success incorporating these extra constraints into a well-behaved linear program --- namely one where our choices of finite-dimensional subspaces $\cW_k$ and discretizations $D=\{(\De_i,l_i)\}$ lead to answers that don't violate other constraints $F_{\De',l'}\geq 0$ for $(\De',l')\notin D$.  It is certainly possible that these difficulties can be circumvented.  However, in this paper, we choose to focus on the information that can be learned from Eq.~(\ref{eq:crossingphiphissuper}).

\begin{figure}
\begin{center}
\begin{psfrags}
\def\PFGstripminus-#1{#1}%
\def\PFGshift(#1,#2)#3{\raisebox{#2}[\height][\depth]{\hbox{%
  \ifdim#1<0pt\kern#1 #3\kern\PFGstripminus#1\else\kern#1 #3\kern-#1\fi}}}%
\providecommand{\PFGstyle}{}%
%
\psfrag{D2d}[cc][cc]{\PFGstyle $\De=2d$}%
\psfrag{D}[bc][bc]{\PFGstyle $\De$}%
\psfrag{d}[cl][cl]{\PFGstyle $d$}%
\psfrag{maxDphidag}[bc][bc]{\PFGstyle $\max(\De_{\Phi^\dag\Phi})$}%
\psfrag{x10251}[tc][tc]{\PFGstyle $1.025$}%
\psfrag{x1051}[tc][tc]{\PFGstyle $1.05$}%
\psfrag{x10751}[tc][tc]{\PFGstyle $1.075$}%
\psfrag{x111}[tc][tc]{\PFGstyle $1.1$}%
\psfrag{x11251}[tc][tc]{\PFGstyle $1.125$}%
\psfrag{x1151}[tc][tc]{\PFGstyle $1.15$}%
\psfrag{x11}[tc][tc]{\PFGstyle $1$}%
\psfrag{y21}[cr][cr]{\PFGstyle $2$}%
\psfrag{y2251}[cr][cr]{\PFGstyle $2.25$}%
\psfrag{y251}[cr][cr]{\PFGstyle $2.5$}%
\psfrag{y2751}[cr][cr]{\PFGstyle $2.75$}%
\psfrag{y31}[cr][cr]{\PFGstyle $3$}%
\psfrag{y3251}[cr][cr]{\PFGstyle $3.25$}%
\psfrag{y351}[cr][cr]{\PFGstyle $3.5$}%
\psfrag{y3751}[cr][cr]{\PFGstyle $3.75$}%
\psfrag{y41}[cr][cr]{\PFGstyle $4$}%
\includegraphics[width=110mm]{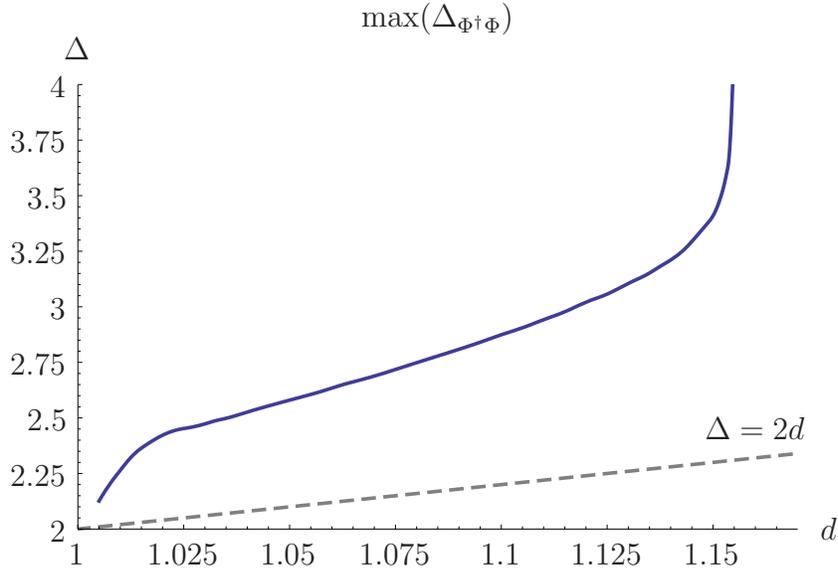}
\end{psfrags}
\end{center}
\caption{An upper bound on the dimension of $\Phi^\dag\Phi$ (the lowest-dimension scalar appearing in the $\Phi\x\Phi^\dag$ OPE), as a function of $d=\dim \Phi$.  Here, we have taken $k=6$.  The bound appears to approach $2$ as $d\rightarrow 1$, as expected.  On the other hand, we do not find a dimension bound for $d\gtrsim 1.16$.  It is possible that pushing the numerics beyond $k=6$ could lead to bounds in this region.}
\label{fig:dimensionbound}
\end{figure}

Figure~\ref{fig:dimensionbound} shows the resulting bound on the dimension of $\Phi^{\dagger} \Phi$ as a function of the dimension $d$ of $\Phi$.  Here we have taken $k=6$, and then for each value of $d$ we scan over values of $\De_0$ (with a spacing of $0.01$) until we find the smallest dimension such that there is a contradiction with unitarity.  As $d \rightarrow 1$, we see that the bound approaches $2$ from above, consistent with the existence of the free theory.  (Bounds very near $d=1$ are computationally intensive to obtain, so we defer very close exploration of this region to future work.)  On the other hand, we see that the bound becomes very weak and shoots off to infinity around $d \sim 1.16$.  For dimensions larger than this value, the resulting bounds on $|\l_{\cO_0}|^2$ become stronger and stronger as $\De_0$ becomes large, but never lead to a violation of unitarity.  We also note that at $k<6$ we do not find a dimension bound at any value of $d$, so that one can only see these bounds when a large number of derivatives are considered.\footnote{As discussed in Appendix~\ref{app:implementation}, $\cW_k$ has dimension $\frac{k(k+1)}{2}$, so that $k=6$ corresponds to $21$ derivatives.  It may be that not all of these derivatives are important for obtaining a dimension bound, and one possible numerical optimization might involve using a subspace of $\cW_6$ other than $\cW_k$ for $k<6$.}  It would be very interesting to see if pushing the numerics further and incorporating even more derivatives could lead to bounds at larger values of $d$.

\begin{figure}
\begin{center}
\begin{psfrags}
\def\PFGstripminus-#1{#1}%
\def\PFGshift(#1,#2)#3{\raisebox{#2}[\height][\depth]{\hbox{%
  \ifdim#1<0pt\kern#1 #3\kern\PFGstripminus#1\else\kern#1 #3\kern-#1\fi}}}%
\providecommand{\PFGstyle}{}%
%
\psfrag{d105}[cc][cc][1][-43]{\PFGstyle $d\!=\!1.05$}%
\psfrag{d11}[cc][cc][1][-40]{\PFGstyle $d\!=\!1.1$}%
\psfrag{d125}[cc][cc][1][-32]{\PFGstyle $d\!=\!1.25$}%
\psfrag{d15}[cc][cc][1][-23]{\PFGstyle $d\!=\!1.5$}%
\psfrag{D}[cl][cl]{\PFGstyle $\De_0$}%
\psfrag{lam}[bc][bc]{\PFGstyle $|\l_{\cO_0}|$}%
\psfrag{maxlforl00}[bc][bc]{\PFGstyle $\max|\l_{\cO_0}|\text{ for }l_0=0$}%
\psfrag{x21}[tc][tc]{\PFGstyle $2$}%
\psfrag{x251}[tc][tc]{\PFGstyle $2.5$}%
\psfrag{x31}[tc][tc]{\PFGstyle $3$}%
\psfrag{x351}[tc][tc]{\PFGstyle $3.5$}%
\psfrag{x41}[tc][tc]{\PFGstyle $4$}%
\psfrag{x451}[tc][tc]{\PFGstyle $4.5$}%
\psfrag{x51}[tc][tc]{\PFGstyle $5$}%
\psfrag{x551}[tc][tc]{\PFGstyle $5.5$}%
\psfrag{y0}[cr][cr]{\PFGstyle $0$}%
\psfrag{y11}[cr][cr]{\PFGstyle $1$}%
\psfrag{y151}[cr][cr]{\PFGstyle $1.5$}%
\psfrag{y21}[cr][cr]{\PFGstyle $2$}%
\psfrag{y251}[cr][cr]{\PFGstyle $2.5$}%
\psfrag{y5}[cr][cr]{\PFGstyle $0.5$}%
\includegraphics[width=120mm]{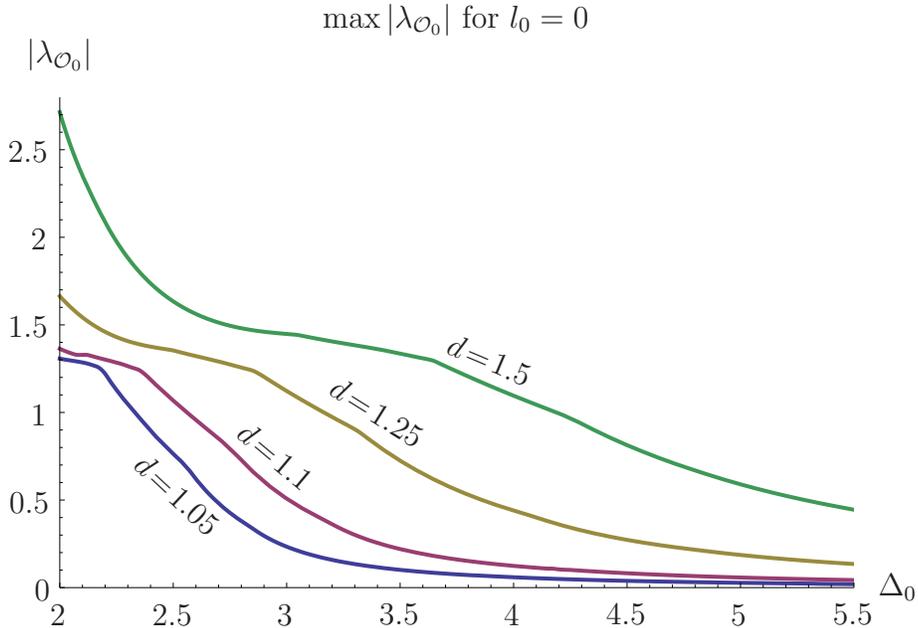}
\end{psfrags}
\end{center}
\caption{An upper bound on the OPE coefficient $|\l_{\cO_0}|$ of a scalar primary $\cO_0$ appearing in $\Phi\x\Phi^\dag$, as a function of $\De_0=\dim\cO_0$, for $d=\dim\Phi=1.05, 1.1, 1.25, 1.5$ (with $k=5$). Here we no longer assume $\cO_0$ is the lowest-dimension scalar in $\Phi\x\Phi^\dag$.}
\label{fig:SUSYscalarOPE}
\end{figure}

We can also consider bounds on the OPE coefficients of operators without making any assumptions about the spectrum.  In this case we simply require that $\a(\cF_{\De,0}) \geq 0$ for all $\De \geq 2$, which is the SUSY unitarity bound for scalar operators with vanishing $U(1)_R$ charge.  In Figure~\ref{fig:SUSYscalarOPE} we show the resulting bounds on $|\l_{\cO_0}|$ for scalar operators appearing in this OPE as a function of their dimension, at various values of $d$.  This is a supersymmetric generalization of the bounds considered in~\cite{Caracciolo:2009bx} in non-supersymmetric theories.  Here we see that the bounds become very strong as $\De_0$ is increased, and appear to approach zero asymptotically.  On the other hand, there are still finite bounds at $\De_0=2$, which tells us that even the coefficients appearing in front of flavor symmetry currents cannot be too large.  We will explore this in more detail in the next subsection.

\subsection{Flavor Currents}
\label{sec:flavorcurrentblocks}

When $\f$ is charged under a flavor symmetry, Ward identities guarantee that flavor currents appear in the OPE $\f\x\f^*$, and thus contribute non-trivially to the conformal block expansion of $\<\f\f^*\f\f^*\>$.  In this subsection, we review the relevant Ward identities for both general CFTs and superconformal theories, and compute these conformal block contributions. In the next section, we present bounds on these quantities.  Although we will eventually specialize to the case of a single operator $\f$, let us first consider a collection of operators $\f_i$ transforming in some general representation under a flavor group $G$.

Suppose $G$ has generators $T^I$ and conserved currents $J^{Ia}$.  The flavor charges in radial quantization are given by integrating the radial component of $J^{Ia}$ over a three-sphere surrounding the origin, $Q^I\equiv -i\int d\Omega\,\hat x\.J^{I}$.  These charges then act non-trivially on our $\f_i$ as $Q^I\f_i(0)= -(T^I)^j_i\f_j(0)$.\footnote{A word about $i$'s and $-1$'s.  The minus sign in the action of $Q^I$ on $\f_i$ ensures that commutators of $Q^I$'s act correctly.  The $-i$ in the definition of $Q^I$ comes from Wick-rotation to Euclidean signature.  This is easiest to see in the usual time slicing, where $\int d^3\vec x J^0\to-i\int d^3\vec x J_E^0$ under $J^0\to -iJ_E^0$.  OPE coefficients that don't involve $\e$-tensors are the same in Euclidean and Lorentzian signature, so we are free to compute them in either signature.}  Comparing this action with our expression for $Q^I$, we see that the $J\x\f$ OPE must take the form
\ben
\label{eq:JphiOPE}
J^{Ia}(x)\f_i(0) &\sim& -\frac i {2\pi^2}(T^I)^j_i\frac{x^a}{x^4} \f_j(0) + \dots,
\een
where ``$\dots$" represents other operators, and we have used that $\mathrm{vol}(S^3)=2\pi^2$.

Suppose the $\f_i$'s and $J^{Ia}$'s are normalized so that
\ben
\label{eq:phinormalization}
\<\f_i(x_1)\f_{\bar\imath}^*(x_2)\> &=& \frac{g_{i\bar\imath}}{x_{12}^{2d}},
\quad \textrm{and}
\\
\label{eq:flavorcurrentnormalization}
\<J^{Ia}(x_1)J^{Jb}(x_2)\> &=& \frac{\tau^{IJ}}{(2\pi)^4}(\ptl^2 \eta^{ab}-\ptl^a\ptl^b)\frac{1}{x_{12}^4}
\ \ =\ \  12\frac{\tau^{IJ}}{(2\pi)^4}\frac{I^{ab}(x_{12})}{x_{12}^6}.
\een
Together, Eqs.~(\ref{eq:JphiOPE}) and (\ref{eq:phinormalization}) give the three-point function
\ben
\<\f_i(x_1) \f_{\bar\jmath}^*(x_2)J^{Ia}(x_3)\> &=& -\frac i {2\pi^2} T^I_{i\bar\jmath}\frac{x_{12}^{2-2d}}{x_{13}^2 x_{23}^2}Z^a,
\een
where $T^I_{i\bar\jmath}\equiv (T^I)_i^jg_{j\bar\jmath}$.  Combining this with Eq.~(\ref{eq:flavorcurrentnormalization}), we find that the conformal block corresponding to an exchange of flavor currents in the $\f\x\f^*$ channel is given by
\ben
\label{eq:bosonicflavorcontribution}
x_{12}^{2d} x_{34}^{2d} \<\f_i\f_{\bar\imath}^*\f_j\f_{\bar\jmath}^*\> &\sim& -\frac 1 3 \tau_{IJ}T^I_{i\bar\imath}T^J_{j\bar\jmath}\,g_{3,1}
\qquad\textrm{(general CFTs),}
\een
where $\tau_{IJ}$ is the inverse of $\tau^{IJ}$.

In superconformal theories, flavor currents $J^{Ia}(x)$ are the $\th\s^a\bar\th$ terms in scalar supermultiplets $J^I(z)$ of dimension $2$.  Comparing Eq.~(\ref{eq:bosonicflavorcontribution}) to the superconformal block Eq.~(\ref{eq:N=1superconformalblock}) with $\De=2$ and $l=0$, we see that flavor supermultiplets contribute to a four-point function of anti-/chiral superconformal primaries as
\ben
x_{12}^{2d} x_{34}^{2d} \<\f_i\f_{\bar\imath}^*\f_j\f_{\bar\jmath}^*\> &\sim& \tau_{IJ}T^I_{i\bar\imath}T^J_{j\bar\jmath}\,\cG_{2,0}
\qquad
\textrm{(SCFTs).}
\een

Although the coefficients $\tau^{IJ}$ are incalculable in general CFTs, in superconformal theories they have a simple expression in terms of the $U(1)_R$ generator~\cite{Anselmi:1997am,Anselmi:1997ys}:
\ben\label{eq:tauSUSY}
\tau^{IJ} &=& -3\Tr(RT^I T^J),
\een
where the trace stands for the coefficient of the $U(1)_R$ anomaly induced by weakly gauging the flavor currents $J^{Ia}$.  For those SCFTs which emerge from a weakly coupled UV theory, this can often be calculated via 't Hooft anomaly matching.

\subsubsection{Flavor Bounds}
\label{sec:flavorbounds}

Consider now a single scalar primary $\f=\f_1$, normalized so that $g_{1\bar 1}=1$.  We can bound the flavor current contribution $\tau_{IJ}T^I_{1\bar 1}T^J_{1\bar 1}$ using the same procedure described in Section~\ref{sec:OPEcoeffbounds}, with slightly modified constraints on the linear functional $\a$.  First, we demand that $\a(F_{\De,l})\geq 0$ (or $\a(\cF_{\De,l})\geq 0$ in the supersymmetric case) for all pairs $(\De,l)$ obeying the relevant unitarity bound.  In general, this is $\De\geq 1$ when $l=0$, and $\De\geq l+2$ otherwise, while in a supersymmetric theory, it is simply $\De\geq l+2$.  We also require $\a(F_{3,1})=1$ in the non-supersymmetric case and $\a(\cF_{2,0})=1$ in the supersymmetric case, since these are the conformal blocks whose coefficients we wish to study.  Note that we are no longer making implicit assumptions about the spectrum of operators appearing in $\f\x\f^*$, so the resulting bounds hold in any unitary CFT with a charged scalar primary.

\begin{figure}
\begin{center}
\begin{psfrags}
\def\PFGstripminus-#1{#1}%
\def\PFGshift(#1,#2)#3{\raisebox{#2}[\height][\depth]{\hbox{%
  \ifdim#1<0pt\kern#1 #3\kern\PFGstripminus#1\else\kern#1 #3\kern-#1\fi}}}%
\providecommand{\PFGstyle}{}%
%
\psfrag{d}[cl][cl]{\PFGstyle $d$}%
\psfrag{maxtaufora}[bc][bc]{\PFGstyle $\max(\tau_{IJ}T^I T^J)\text{ for a charged scalar}$}%
\psfrag{tau}[bc][bc]{\PFGstyle $\tau_{IJ}T^I T^J$}%
\psfrag{x111}[tc][tc]{\PFGstyle $1.1$}%
\psfrag{x11}[tc][tc]{\PFGstyle $1$}%
\psfrag{x121}[tc][tc]{\PFGstyle $1.2$}%
\psfrag{x131}[tc][tc]{\PFGstyle $1.3$}%
\psfrag{x141}[tc][tc]{\PFGstyle $1.4$}%
\psfrag{x151}[tc][tc]{\PFGstyle $1.5$}%
\psfrag{y0}[cr][cr]{\PFGstyle $0$}%
\psfrag{y12}[cr][cr]{\PFGstyle $10$}%
\psfrag{y152}[cr][cr]{\PFGstyle $15$}%
\psfrag{y22}[cr][cr]{\PFGstyle $20$}%
\psfrag{y252}[cr][cr]{\PFGstyle $25$}%
\psfrag{y32}[cr][cr]{\PFGstyle $30$}%
\psfrag{y352}[cr][cr]{\PFGstyle $35$}%
\psfrag{y42}[cr][cr]{\PFGstyle $40$}%
\psfrag{y51}[cr][cr]{\PFGstyle $5$}%
\includegraphics[width=100mm]{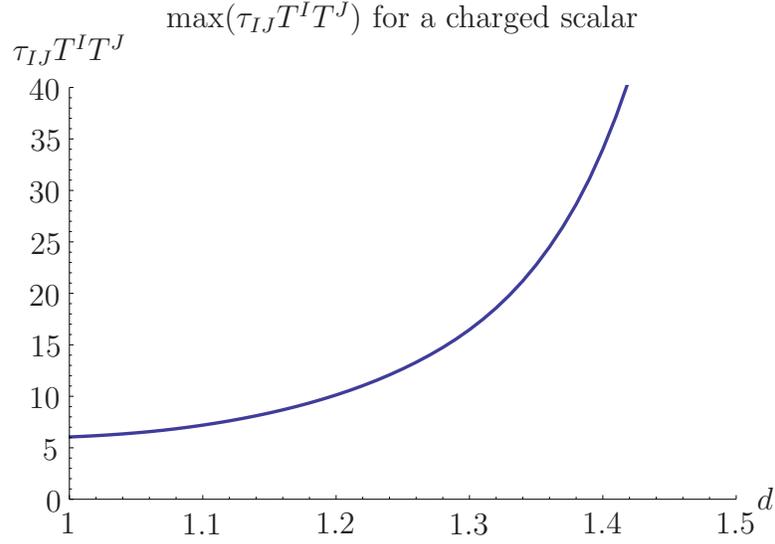}
\end{psfrags}
\end{center}
\caption{An upper bound on $\tau_{IJ}T^{I}_{1\bar 1}T^J_{1\bar1}$ which is $-3$ times the flavor current coefficient in the OPE $\f\x\f^*$ of a complex scalar with its conjugate, as a function of $d=\dim \f$. Here $k=5$.}
\label{fig:chargedFlavorPlot}
\end{figure}

\begin{figure}
\begin{center}
\begin{psfrags}
\def\PFGstripminus-#1{#1}%
\def\PFGshift(#1,#2)#3{\raisebox{#2}[\height][\depth]{\hbox{%
  \ifdim#1<0pt\kern#1 #3\kern\PFGstripminus#1\else\kern#1 #3\kern-#1\fi}}}%
\providecommand{\PFGstyle}{}%
%
\psfrag{d}[cl][cl]{\PFGstyle $d$}%
\psfrag{maxtaufora}[bc][bc]{\PFGstyle $\max(\tau_{IJ}T^I T^J)\text{ for a chiral primary}$}%
\psfrag{tau}[bc][bc]{\PFGstyle $\tau_{IJ}T^I T^J$}%
\psfrag{x111}[tc][tc]{\PFGstyle $1.1$}%
\psfrag{x11}[tc][tc]{\PFGstyle $1$}%
\psfrag{x121}[tc][tc]{\PFGstyle $1.2$}%
\psfrag{x131}[tc][tc]{\PFGstyle $1.3$}%
\psfrag{x141}[tc][tc]{\PFGstyle $1.4$}%
\psfrag{x151}[tc][tc]{\PFGstyle $1.5$}%
\psfrag{x161}[tc][tc]{\PFGstyle $1.6$}%
\psfrag{y0}[cr][cr]{\PFGstyle $0$}%
\psfrag{y12}[cr][cr]{\PFGstyle $10$}%
\psfrag{y21}[cr][cr]{\PFGstyle $2$}%
\psfrag{y41}[cr][cr]{\PFGstyle $4$}%
\psfrag{y61}[cr][cr]{\PFGstyle $6$}%
\psfrag{y81}[cr][cr]{\PFGstyle $8$}%
\includegraphics[width=100mm]{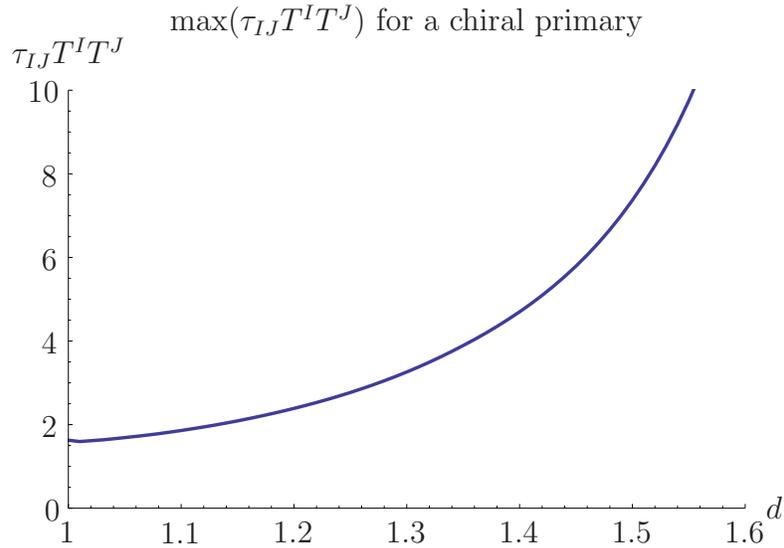}
\end{psfrags}
\end{center}
\caption{An upper bound on the flavor current coefficient $\tau_{IJ}T^{I}_{1\bar 1}T^J_{1\bar1}$ appearing in the OPE $\f\x\f^*$ of a chiral primary with its conjugate in an $\cN=1$ SCFT, as a function of $d=\dim\f$. Here $k=5$.}
\label{fig:SUSYFlavorPlot}
\end{figure}

An upper bound on $\tau_{IJ}T^I_{1\bar 1}T^J_{1\bar 1}$ as a function of $d=\dim\f_1$ is shown in Figure~\ref{fig:chargedFlavorPlot} for a general CFT, and Figure~\ref{fig:SUSYFlavorPlot} for a superconformal theory.  Both bounds are strongest when $d$ is near 1, and become weaker as $d$ increases.  The supersymmetric bound is most stringent, requiring $\tau_{IJ}T^I_{1\bar 1}T^J_{1\bar 1}\lesssim 1.6$ when $d\aeq 1$.

Let us pause for a moment to appreciate the non-trivial nature of these bounds.  If we for example consider a global $U(1)$ symmetry with charges $Q_i$ in an asymptotically free superconformal theory, then using Eq.~(\ref{eq:tauSUSY}) we are placing an upper bound on the quantity
\ben
\frac{Q^2}{-3 \sum_{i} (R_i-1)Q_i^2},
\een
where the sum runs over chiral superfields in the UV description, and we are considering a gauge-invariant operator with charge $Q$.  First note that this quantity does not depend on the overall normalization of the $U(1)$ charges, which is unphysical.  Our bound immediately tells us that one cannot have a global $U(1)$ symmetry that acts only on fields that have $R$ very close to $1$.  In addition, in principle one could imagine a cancellation between terms in the denominator, since some fields may have $R$ smaller than $1$ and some may have $R$ greater than $1$.  Our bound also tells us that an arbitrary cancellation between terms is not possible.  

These bounds are also potentially interesting in light of the AdS/CFT correspondence~\cite{Maldacena:1997re,Gubser:1998bc,Witten:1998qj}.  In this case $\tau^{I J}$ is directly mapped to the size of the coupling constants for the corresponding bulk gauge fields. In AdS$_5$ this mapping is given by~\cite{Freedman:1998tz}
\ben
\tau^{I J} &=& 8 \pi^2 L (g^{-2})^{I J},
\een
where $L$ is the AdS length scale and the gauge coupling $(g^{-2})^{I J}$ appears in the action as
\ben
S_{AdS} = \int d^5 x \sqrt{-g}\left[\frac14 (g^{-2})^{I J} F_{I}^{\mu\nu} F_{J \mu\nu}+\dots\right] .
\een
Our bound tells us that there is a fundamental obstruction to making the gauge couplings arbitrarily large in the presence of charged scalar bulk excitations corresponding to operators with dimension close to $1$.  It would be very interesting to explore this connection further in a controlled setting, and to see if there is any kind of bulk reasoning that could give rise to this bound.

\subsection{The Stress Tensor}

The stress tensor $T^{ab}$ also makes a non-trivial contribution to conformal block expansions.  Let us now review the relevant Ward identities, and compute the coefficient of the stress tensor conformal block in the supersymmetric and non-supersymmetric case.  In the following subsection we will present bounds on these contributions.

The dilatation operator in radial quantization is given by $D=(-i)^2\int d\Omega\,\hat x_a x_b T^{ab}$, where the integral is over a three-sphere surrounding the origin.  Requiring the action $D\f(0)=d\f(0)$ then determines the OPE
\ben
\label{eq:stressOPE}
T^{ab}(x)\f(0) &\sim& -\frac{4d}{3}\frac 1 {2\pi^2}\p{x^a x^b-\frac 1 4 \eta^{ab}x^2}\frac 1 {x^6}\f(0)+\dots
\een
The stress tensor is conventionally normalized as
\ben
\label{eq:stressNorm}
\<T^{ab}(x)T^{cd}(0)\> &=& \frac{40 c}{\pi^4}\frac{I^{ac}(x)I^{bd}(x)} {x^8},
\een
where we implicitly symmetrize and subtract traces in each pair of indices $a,b$ and $c,d$ on the right-hand side.  The coefficient $c$ is the central charge, which appears for example in the trace anomaly $\<T_a^a\>=\frac c {16\pi^2}(\mathrm{Weyl})^2-\frac a {16\pi^2}(\mathrm{Euler})$ of the theory on a curved background.  In our conventions, a free real scalar has $c=\frac 1 {120}$ while a free Weyl fermion has $c=\frac 1 {40}$.  Just as with the flavor current normalization $\tau^{IJ}$, there is an explicit formula for $c$ in superconformal theories,
\ben\label{eq:cSUSY}
c &=& \frac 1 {32} (9\Tr R^3-5 \Tr R),
\een
where $R$ is the $U(1)_R$ generator, and the traces stand for anomaly coefficients.  For a free chiral superfield ($R=2/3$), the above equation yields $c=\frac 1 {24}$.

Now combining Eqs.~(\ref{eq:stressOPE}) and (\ref{eq:stressNorm}), we obtain the stress tensor conformal block contribution
\ben
\label{eq:stressconformalblock}
x_{12}^{2d} x_{34}^{2d} \<\f \f^*\f\f^*\> &\sim& \frac{d^2}{90 c}\,g_{4,2},
\qquad
(\textrm{general CFTs})
\een
where we've assumed that $\f$ is normalized to have $\<\f(x)\f^*(0)\>=x^{-2d}$.

In superconformal theories, the stress tensor is the $\th\s^a\bar\th$ component of the supercurrent $\cJ^a(z)$~\cite{Ferrara:1974pz}, a supermultiplet with dimension $3$ and spin $1$.  Comparing with the superconformal block Eq.~(\ref{eq:N=1superconformalblock}), we see that the supercurrent contribution is
\ben
\label{eq:supercurrentconformalblock}
x_{12}^{2d} x_{34}^{2d}  \<\f \f^*\f\f^*\> &\sim& -\frac{d^2}{36 c}\,\cG_{3,1}
\qquad
(\textrm{SCFTs})
\een
Note that the lowest component of $\cJ^a$ is the $U(1)_R$ current $J_R^a$, which has a conformal block contribution dictated by Eq.~(\ref{eq:bosonicflavorcontribution}) with $T^R=\frac 2 3 d$.  Comparing with Eq.~(\ref{eq:supercurrentconformalblock}), we find $\tau^{RR}=\frac{16 c}{3}$, which is indeed correct (see for example \cite{Barnes:2005bm}).

\subsubsection{Central Charge Bounds}

We can now produce bounds on the central charge $c$ using the same procedure as for flavor currents in Section~\ref{sec:flavorbounds}, but with the equality constraints modified to $\a(F_{4,2})=1$ in the non-supersymmetric case and $\a(\cF_{3,1})=1$ in the supersymmetric case.  Note that the coefficients in Eqs.~(\ref{eq:stressconformalblock}) and (\ref{eq:supercurrentconformalblock}) are inversely proportional to $c$, so that an upper bound on conformal block coefficients implies a {\it lower} bound on the central charge $c\geq f_c(d)$, as a function of $d=\dim\f$.

\begin{figure}
\begin{center}
\begin{psfrags}
\def\PFGstripminus-#1{#1}%
\def\PFGshift(#1,#2)#3{\raisebox{#2}[\height][\depth]{\hbox{%
  \ifdim#1<0pt\kern#1 #3\kern\PFGstripminus#1\else\kern#1 #3\kern-#1\fi}}}%
\providecommand{\PFGstyle}{}%
%
\psfrag{c}[bc][bc]{\PFGstyle $c$}%
\psfrag{d}[cl][cl]{\PFGstyle $d$}%
\psfrag{freescalar}[cc][cc]{\PFGstyle $\text{free scalar}$}%
\psfrag{k3}[cc][cc][1][-38]{\PFGstyle $\!\!k\!=\!3$}%
\psfrag{k4}[cc][cc][1][-44]{\PFGstyle $\!k\!=\!4$}%
\psfrag{k5}[cc][cc][1][-37]{\PFGstyle $k\!=\!5$}%
\psfrag{k6}[cc][cc][1][-45]{\PFGstyle $k\!=\!6$}%
\psfrag{mincforgen}[bc][bc]{\PFGstyle $\min(c)\text{ for general CFT}$}%
\psfrag{x11}[tc][tc]{\PFGstyle $1$}%
\psfrag{x121}[tc][tc]{\PFGstyle $1.2$}%
\psfrag{x141}[tc][tc]{\PFGstyle $1.4$}%
\psfrag{x161}[tc][tc]{\PFGstyle $1.6$}%
\psfrag{x181}[tc][tc]{\PFGstyle $1.8$}%
\psfrag{x21}[tc][tc]{\PFGstyle $2$}%
\psfrag{y0}[cr][cr]{\PFGstyle $0$}%
\psfrag{y1m1}[cr][cr]{\PFGstyle $0.01$}%
\psfrag{y2m2}[cr][cr]{\PFGstyle $0.002$}%
\psfrag{y4m2}[cr][cr]{\PFGstyle $0.004$}%
\psfrag{y6m2}[cr][cr]{\PFGstyle $0.006$}%
\psfrag{y8m2}[cr][cr]{\PFGstyle $0.008$}%
\includegraphics[width=120mm]{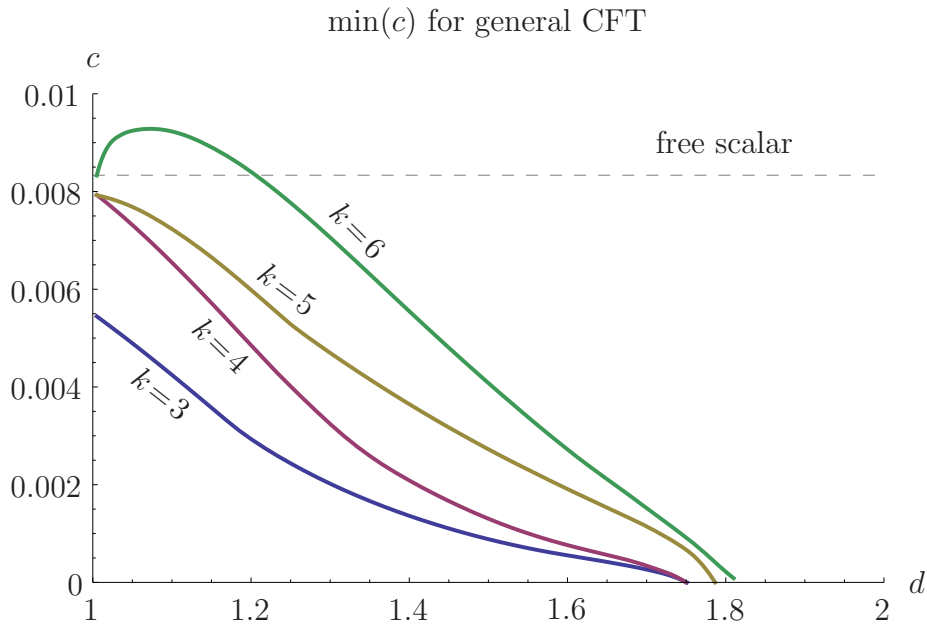}
\end{psfrags}
\end{center}
\caption{A lower bound on the central charge $c$ in a general CFT containing a real scalar primary $\f$, as a function of $d=\dim\f$.  Here, we show the optimal bound for various values of $k$, corresponding to restrictions of the search to different finite-dimensional subspaces $\cW_k\subset\cV^*$ (Eq.~\ref{eq:Wksuspaces}).  As $k$ increases, the bound gets stronger, though it becomes more computationally intensive to obtain.  The bounds possibly converge to the free scalar value $c=1/120$ as $d \rightarrow 1$.}
\label{fig:realcentralcharge}
\end{figure}

We plot this bound for the case of a real scalar in Figure~\ref{fig:realcentralcharge}.  We have included curves for different values of $k$ (indexing the size of our finite-dimensional subspaces $\cW_k$) to show how the bound gets stronger as we widen the search space $\cS\cap \cW_k$.  In particular, as $k$ increases, the series of bounds $c\geq f^{(k)}_c(d)$ appears to approach $c\geq c_\textrm{free scalar}$ at $d=1$.  Recall that precisely at $d=1$, the $\f$ operator is free and decouples from the rest of the theory, contributing exactly $c_\textrm{free scalar}$ to $c$.  We conjecture that  $f^{(k)}_c(1)\to c_\textrm{free scalar}$ as $k\to \oo$, namely that the optimal bound at $d=1$ can be achieved with these methods.

\begin{figure}
\begin{center}
\begin{psfrags}
\def\PFGstripminus-#1{#1}%
\def\PFGshift(#1,#2)#3{\raisebox{#2}[\height][\depth]{\hbox{%
  \ifdim#1<0pt\kern#1 #3\kern\PFGstripminus#1\else\kern#1 #3\kern-#1\fi}}}%
\providecommand{\PFGstyle}{}%
%
\psfrag{cmin}[bc][bc]{\PFGstyle $\min(c)$}%
\psfrag{d}[cl][cl]{\PFGstyle $d$}%
\psfrag{freechiral}[cc][cc]{\PFGstyle $\text{free chiral superfield}$}%
\psfrag{mincforSCF}[bc][bc]{\PFGstyle $\min(c)\text{ for SCFT}$}%
\psfrag{x11}[tc][tc]{\PFGstyle $1$}%
\psfrag{x121}[tc][tc]{\PFGstyle $1.2$}%
\psfrag{x141}[tc][tc]{\PFGstyle $1.4$}%
\psfrag{x161}[tc][tc]{\PFGstyle $1.6$}%
\psfrag{x181}[tc][tc]{\PFGstyle $1.8$}%
\psfrag{x21}[tc][tc]{\PFGstyle $2$}%
\psfrag{y0}[cr][cr]{\PFGstyle $0$}%
\psfrag{y1m1}[cr][cr]{\PFGstyle $0.01$}%
\psfrag{y2m1}[cr][cr]{\PFGstyle $0.02$}%
\psfrag{y3m1}[cr][cr]{\PFGstyle $0.03$}%
\psfrag{y4m1}[cr][cr]{\PFGstyle $0.04$}%
\includegraphics[width=100mm]{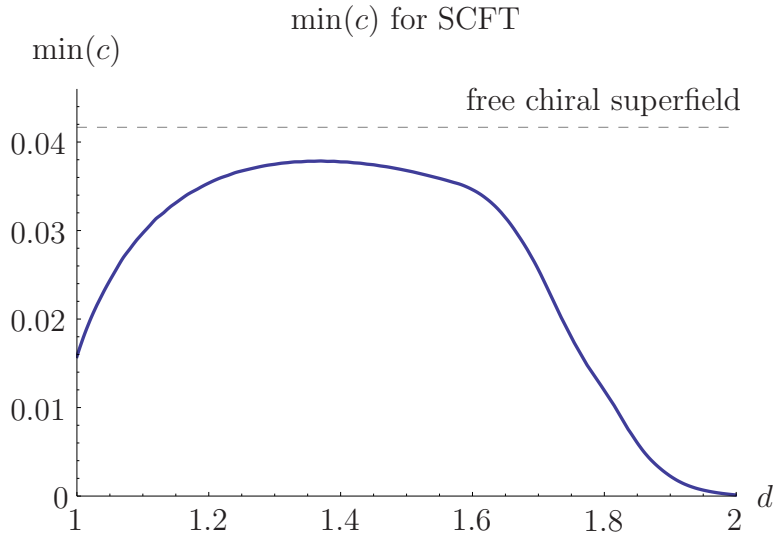}
\end{psfrags}
\end{center}
\caption{A lower bound on the central charge $c$ of a superconformal theory containing a scalar chiral primary $\f$, as a function of $d=\dim\f$.  Here we have taken $k=5$.  Note that the assumption of supersymmetry allows us to strengthen the bound significantly from the real scalar case (Figure~\ref{fig:realcentralcharge}).  However, the bound does not appear to approach the free chiral superfield value $c=1/24$ near $d=1$.  This is likely because we are only using the partial crossing relation Eq.~(\ref{eq:crossingphiphis}) instead of the full information in Eq.~(\ref{eq:crossingphiphi}).}
\label{fig:SUSYcentralchargeplot}
\end{figure}

In Figure~\ref{fig:SUSYcentralchargeplot}, we plot a lower bound on the central charge $c\geq f_c^\textrm{SUSY}(d)$, in any superconformal theory containing a scalar chiral primary of dimension $d$.  Incorporating constraints from superconformal symmetry into the crossing relations certainly gives a stronger bound than in the case of a real scalar.   However, we do not have $f^\textrm{SUSY}_c(1)\sim c_\textrm{free chiral superfield}$, possibly reflecting the fact that we are using only the partial crossing relation Eq.~(\ref{eq:crossingphiphissuper}).

Unfortunately these central charge bounds are not ``additive."  That is, they are not stronger in the presence of multiple degrees of freedom, unless those degrees of freedom are completely decoupled from one another.  Consider, for instance, a CFT with $n$ real scalars transforming in the fundamental of an $\SO(n)$ flavor symmetry.  If these scalars are decoupled, each with central charge $c$, then we can safely write $T^{ab}_\textrm{full theory}=\sum_iT^{ab}_i$, where $T^{ab}_i$ is the stress tensor in the $i$-th decoupled sector, and compute $c_\textrm{full theory}=n c$.  In this case, a lower bound on $c$ translates trivially into a much stronger lower bound on $c_\textrm{full theory}$.  However, now suppose our scalars are weakly interacting.  We no longer have separate conserved stress tensors $T_i^{ab}$, but rather a single stress tensor $T^{ab}$ which is a singlet under $\SO(n)$, along with a non-conserved spin-2 operator $T^{ab}_\textrm{fake}$ transforming in a traceless symmetric tensor of $\SO(n)$.  We can no longer say what the dimension of $T^{ab}_\textrm{fake}$ is, since it's no longer protected by a conservation law, and consequently we cannot straightforwardly include it in a linear program.

While our bounds on $c$ are perhaps somewhat weak, they are still highly non-trivial.  In superconformal theories, for example, there is no a priori reason to think that different contributions to the central charge in Eq.~(\ref{eq:cSUSY}) cannot cancel each other to a high degree.  However, our bound says that this is not possible if the theory contains a scalar chiral primary operator of low dimension.  

In the context of AdS/CFT, $c$ can be related to the bulk Planck scale $M_P$ and AdS length scale $L$ as $c \sim \pi^2 L^3 M_P^3 $~\cite{Henningson:1998gx}.  Our bound then suggests that there is a fundamental obstruction to making quantum gravity on AdS$_5$ arbitrarily strong in the presence of bulk scalar excitations corresponding to operators with dimension close to $1$.  It would be very interesting to make this more precise, and to understand the origin of the bound from the bulk perspective.

Finally, we note that a somewhat similar bound on the central charge was derived in~\cite{Hellerman:2009bu} in the context of 2D conformal field theories.  There, an inequality relating $c$ to the dimension of the lowest-dimension primary operator was derived through the use of unitarity and modular invariance, the latter of which is not available in 4 dimensions.\footnote{Or rather, modular invariance of the partition function on $T^4\cong T^3\x S^1$ doesn't lead to simple statements about the spectrum of operators, which is given by quantizing the theory on $S^3\!\x\!\textrm{time}$ and not $T^3\!\x\!\textrm{time}$.}  It would be interesting to derive bounds on $c$ in 2-dimensional theories using the present techniques and see how the results compare to those of~\cite{Hellerman:2009bu}.

\section{Comparison to Known Theories}\label{sec:examples}
Since the SUSY flavor and central charge bounds derived in the previous section apply to quantities that are computable via 't Hooft anomaly matching using Eqs.~(\ref{eq:tauSUSY}) and~(\ref{eq:cSUSY}), we can check whether they are satisfied in theories believed to flow to superconformal fixed points.  Doing so requires knowing how the $U(1)_R$ subgroup of the superconformal group acts on our theory.  In simple cases this action is uniquely determined by symmetry considerations, but in general it must be determined using $a$-maximization~\cite{Intriligator:2003jj}.  This requires knowing the full set of IR flavor symmetries that can mix with $U(1)_R$, and in many cases accidental symmetries can arise that are not apparent from the UV description of the theory (for a nice discussion see~\cite{Leigh:1996ds}).  In practice, one can sometimes identify the emergence of such accidental symmetries through apparent violations of the unitarity bound for the dimension of chiral primary operators, or by using a Seiberg dual description~\cite{Seiberg:1994pq} if one is available.  For a number of examples of such analyses, see e.g.~\cite{Kutasov:2003iy,Intriligator:2003mi,Kutasov:2003ux,Barnes:2004jj,Csaki:2004uj,Barnes:2005zn,Intriligator:2005if,Kawano:2005nc,Kawano:2007rz,Shapere:2008un,Poland:2009px}.  In principle, apparent violation of our bounds could provide additional evidence for the emergence of such accidental symmetries, or even the absence of a superconformal fixed point.

In this section, we will develop some intuition for the strength of our bounds by applying them to some simple superconformal theories, namely $\SU(N_c)$ SQCD in the conformal window, and SQCD with an adjoint $X$ and superpotential $\Tr X^3$.  In both cases, we find that the bounds are most interesting at small $N_c$, although they are easily satisfied for all $N_c\geq 2$ throughout the conformal window.  While there are of course many other theories that can be checked, we will leave a more comprehensive survey of $\cN=1$ theories to future work.

\subsection{SQCD}
Let us start by considering $SU(N_c)$ SUSY QCD with $N_f$ vector-like flavors $Q^i$ and ${\tl Q}_{\tl \imath}$.  For $\frac{3}{2} N_c < N_f < 3 N_c$ the theory is believed to flow to an interacting conformal fixed point~\cite{Seiberg:1994pq}.  The anomaly-free global symmetries are $SU(N_f)_L \times SU(N_f)_R \times U(1)_B \times U(1)_R$, with $Q^i$ transforming as $(N_f, 1,1,1 - \frac{N_f}{N_c})$ and ${\tl Q}_{\tl \imath}$ transforming as $(1,\bar{N}_f,-1,1-\frac{N_f}{N_c})$.   The ring of gauge-invariant chiral primary operators is generated by mesons $M^i_{\tl \jmath} = Q^i \tl Q_{\tl \jmath}$ and baryons $B^{i_1 \dots i_{N_c}} =  Q^{i_1} \dots Q^{i_{N_c}}$ and ${\tl B}_{{\tl \imath}_1 \dots {\tl \imath}_{N_c}} = {\tl Q}_{\tl \imath_1} \dots {\tl Q}_{\tl \imath_{N_c}}$.

Mesons have dimension $d_M = 3(1-N_c/N_f)$ which can be close to $1$ near the lower end of the conformal window $N_f \sim \frac{3}{2} N_c$.  Thus, they are good candidates for operators on which to check our bounds.  Concretely, we may pick out a single component $M^1_1$, and consider the constraints that crossing symmetry imposes on the superconformal block decomposition of the four-point function $\< M^1_1 M_1^{1\dagger} M_{1}^1 M_{1}^{1\dagger} \>$.  In particular, the $M_{1}^1 \times M_{1}^{1\dagger}$ OPE contains both $SU(N_f)_L$ and $SU(N_f)_R$ flavor currents, along with the supercurrent $\cJ^a$.

Let us focus on $\SU(N_f)_L$ and first compute $\tau^{IJ}$.  We can work in the UV using 't Hooft anomaly matching.  The fermions contained in $Q$ have $R$-charge $-\frac {N_c}{N_f}$, so we find
\ben
\tau^{I J} = -3\Tr(RT^I T^J)= \frac{3 N_c^2}{2N_f}\de^{I J},
\een
where the generators $T^I$ of $\SU(N_f)_L$ have the usual normalization $\Tr(T^I T^J)=\frac 1 2 \de^{I J}$ in the fundamental representation, and the extra $N_c$ factor comes from summing over colors.  Thus, we obtain
\ben
\tau_{I J} T^I_{1 1} T^J_{1 1} = \frac{2N_f}{3N_c^2}\delta_{IJ}T^I_{1 1}T^J_{1 1} = \frac{N_f-1}{3N_c^2},
\een
where we have used the contraction $\de_{I J} T^{I}_{i\bar\imath} T^{J}_{j\bar\jmath} = \frac 1 2(\de_{i\bar\jmath}\de_{j\bar\imath}-\frac 1 {N_f}\de_{i\bar \imath}\de_{j\bar \jmath})$.
Now, the meson $M_1^1$ also gets an equal contribution from $\SU(N_f)_R$, so the total contribution from flavor currents is  $2(N_f-1)/3N_c^2$.  Note that this scales as $\sim 1/N_c$ for large $N_c$ and fixed $N_f/N_c$, so the bound is mainly interesting for small $N_c$ theories.   However, it is readily verified that all values of $N_f$ and $N_c$ within the conformal window satisfy the bound given in Fig.~\ref{fig:SUSYFlavorPlot}.  For example, taking $N_c=2$ and $N_f=4$ we have $d_M = 1.5$ and a coefficient of  $.5$, whereas the bound tells us that the coefficient cannot be larger than $\sim 6$.  Similarly, taking $N_c=3$ and $N_f=5$ we have $d_M = 1.2$ and a coefficient of $.3$, whereas the bound is $\sim 2$.   Thus, while these theories are a factor of a few away, they do not come very close to saturating the bound.

Finally, let us compare the central charge to the bound given in Fig.~\ref{fig:SUSYcentralchargeplot}.  We can calculate
\ben
c = \frac{1}{32}\left( 9\Tr R^3 - 5 \Tr R \right)  = \frac{1}{16} \left( 7 N_c^2 - 9 \frac{N_c^4}{N_f^2} -2\right).
\een
Note that this grows like $\sim N_c^2$ for fixed $N_f/N_c$, so the bound is again most likely to be interesting for small $N_c$ theories.  However, it is also interesting that there are contributions to $c$ with opposite signs, so in principle there could have been a cancellation.  This occurs at $N_f \sim 3 N_c^2 / \sqrt{7 N_c^2-2}$, which is always outside the conformal window.  On the other hand, for all values of $N_c$ and $N_f$ inside the conformal window $c$ is greater than $1$, and hence easily satisfies the bound in Fig.~\ref{fig:SUSYcentralchargeplot}.

\subsection{SQCD with an Adjoint}
Let us next consider $SU(N_c)$ SUSY QCD with $N_f$ flavors and an adjoint $X$.  For simplicity, we focus on the theory with superpotential $W = \Tr X^{3}$, which was studied in detail in~\cite{Kutasov:1995ve}.\footnote{Note that one can straightforwardly generalize this discussion to the case of $W=\Tr X^{k+1}$ or a vanishing superpotential using the results of~\cite{Kutasov:1995np,Kutasov:1995ss,Kutasov:2003iy}.}  This theory is believed to flow to an interacting fixed point for $\frac{2}{3} N_c < N_f < 2 N_c$.  The anomaly-free global symmetries are $SU(N_f)_L \times SU(N_f)_R \times U(1)_B \times U(1)_R$, with $Q^i$ transforming as $(N_f,1,1,1-\frac{2}{3} \frac{N_c}{N_f})$, ${\tl Q}_{\tl \imath}$ transforming as $(1,\bar{N}_f,-1,1-\frac{2}{3} \frac{N_c}{N_f})$, and $X$ transforming as $(1,1,0,\frac{2}{3})$.  Here we will focus on the chiral primary ``meson" operators $M^i_{\tl \jmath} = Q^i {\tl Q}_{\tl \jmath}$ and $N^i_{\tl \jmath} = Q^i X {\tl Q}_{\tl \jmath}$.  The theory also contains the chiral operator $\Tr X^2$ as well as baryons built out of products of $Q^i$ and $X Q^i$, and anti-baryons built out of products of ${\tl Q}_{\tl \imath}$ and $X {\tl Q}_{\tl \imath}$.  Note that when $\frac{2}{3} N_c < N_f < N_c$, $M^i_{\tl \jmath}$ possesses a $U(1)_R$ charge that appears to violate the unitarity bound, and the interpretation in this case is that this operator has decoupled from the theory and become a free field.  
  
Let us begin with the case of $N_f > N_c$. The dimension of $M^i_{\tl \jmath}$ is given by $d_{M} = 3 - \frac{2N_c}{N_f} $ and approaches $1$ for $N_f \sim N_c$.  We will consider the bounds arising from crossing symmetry of the four-point function $\<M_1^1 M_1^{1\dagger} M_1^1 M_1^{1\dagger}\>$.  Each flavor group $\SU(N_f)_{L,R}$ has
\ben
\tau^{IJ} &=& \frac{N_c^2}{N_f}\de^{IJ}.
\een
Since $M$ is a flavor bifundamental, the flavor current conformal block contribution is
\ben
(\tau_{I J} T^I_{1 1} T^J_{1 1})_L+(\tau_{I J} T^I_{1 1} T^J_{1 1})_R &=& \frac{N_f-1}{N_c^2}.
\een
Again this scales as $\sim 1/N_c$, and there is no violation of our bound for any choice of $N_c<N_f<2N_c$. Further, when $N_f>N_c$ the central charge is given by
\ben
c &=& \frac{1}{24}\left(9N_c^2 - 4\frac{N_c^4}{N_f^2}-4\right).
\een
Again, for $N_c < N_f < 2N_c$ we always have $c>1$ and are unable to approach the bound.    

Next let us consider the range $\frac{2}{3} N_c < N_f < N_c$.  In this case the meson $M$ becomes a free field and decouples from the rest of the theory.  In the dual magnetic description this is simply described by the superpotential coupling involving $M$ flowing to zero rather than a fixed point value~\cite{Kutasov:1995ve}. In the present description we may equivalently describe this situation by adding to the theory a superpotential $W_{LM} = L_i^{\tl \jmath} (M^i_{\tl \jmath} - Q^i {\tl Q}_{\tl \jmath})$, containing new gauge-singlet fields $L$ and $M$~\cite{Barnes:2004jj}.  When $N_f > N_c$, $L$ and $M$ are massive and can simply be integrated out, with the $L$ equation of motion setting $M=Q\tl Q$ in the chiral ring.  However, when $N_f < N_c$ the ``mass term" $L M$ flows to zero and $M$ is no longer interacting.  We are left with a single new interacting field $L$ whose equation of motion now sets $Q \tl{Q}$ to zero in the chiral ring, thus avoiding the unitarity constraint.  One must then include both $L$ and $M$ when computing $\tau$ and $c$ via anomaly matching.  Of course, we already know that the central charge is at least as large as the contribution from $M$, so we cannot learn anything new from this bound.  Additionally, since $M$ has decoupled from the theory we can take it to transform under separate flavor symmetries as compared to the interacting sector. 

Now we will investigate the flavor current constraints imposed by crossing symmetry of the four-point function $\<N_1^1 N_1^{1\dagger} N_1^1 N_1^{1\dagger}\>$ in this regime.  For both flavor groups $SU(N_f)_{L,R}$ (which no longer act on $M$, but do act on $L$), we have
\ben
\tau^{IJ} &=& \left(\frac{N_c^2}{N_f} + \frac{3}{2} N_f - 2 N_c\right)\de^{IJ},
\een
from which we obtain
\ben
(\tau_{IJ} T^I_{1 1} T^{J}_{1 1})_L+(\tau_{IJ} T^I_{1 1} T^{J}_{1 1})_R &=& \frac{2N_f-2}{2 N_c^2 + 3 N_f^2 - 4 N_c N_f}.
\een
One can then verify that for all $\frac{2}{3} N_c < N_f < N_c$ the bound of Fig.~\ref{fig:SUSYFlavorPlot} is satisfied.

\section{Conclusions}\label{sec:concl}

Let us point out some possible directions for future research.  First, the bounds obtained in this work can of course be improved with more refined numerical methods.  In the case of operators transforming under global symmetries, it also seems possible that additional crossing constraints can be incorporated that were not utilized in the present study.  
It would be interesting to see if doing so could lead to even stronger bounds on the dimensions of non-chiral operators in superconformal theories, so that one could start probing more phenomenologically interesting scenarios such as those of~\cite{Roy:2007nz,Murayama:2007ge}.  Another interesting application is to see if one can bound the lowest-dimension $SU(2)$-singlet operator in conformal technicolor models, as was extensively discussed in~\cite{Rattazzi:2008pe}.  In an ideal world, by incorporating the full set of crossing constraints one could perhaps obtain bounds that scale with the size of global symmetry representations.  For example, the central charge $c$ should roughly reflect the number of degrees of freedom of a theory, so the presence of large flavor representations should signal larger $c$.\footnote{Actually, the central charge $a$ generally appears to be a better measure of the number of degrees of freedom \cite{Anselmi:1997am,Anselmi:1997ys,Intriligator:2003jj}. However, $c$ is constrained by $c\geq \frac 2 3 a$ in supersymmetric theories and $c\geq \frac{18}{31}a$ in general \cite{Hofman:2008ar}, so $c$ is large whenever $a$ is large.} It would be nice to derive a bound that supports this intuition.

Another goal is to try to find $\cN=1$ SCFTs that come closer to saturating the bounds on $c$ and $\tau_{I J} T^I T^J$.  If any violation of these bounds could be found, it would be evidence for the emergence of new accidental symmetries or perhaps the absence of a conformal fixed point altogether.  It may also be possible to extend these results to $\cN=2$ theories where one could obtain even stronger bounds. 

Similar studies in different numbers of dimensions are also feasable.  The extension to other even dimensions should be completely straightforward since the conformal blocks are known, and it would for example be interesting to see what kind of central charge bounds can be obtained in 2D using the present methods.  While closed-form expressions for the conformal blocks in odd dimensions are not currently available, it seems likely that one could still use recursion relations (as described in Appendix~\ref{app:implementation}) to efficiently evaluate conformal blocks and their derivatives.  Bounds obtained in 3D might then be relevant for condensed matter systems.

Finally, it would be fascinating to better understand the interpretation of these bounds in the context of the AdS/CFT correspondence.  They suggest that there should be a fundamental limit to the strength of gravitational and gauge forces in the presence of light bulk excitations in AdS$_5$.  Since our bounds are most interesting for small $N$ theories, it seems likely that one will have to go to a highly quantum regime in order to see these effects.  Nevertheless, it would be interesting to see if there is any simple bulk reasoning that could shed light on the origin of these bounds.  One might then hope that thinking about these issues could lead to a deeper understanding of the nature of quantum gravity.

\section*{Acknowledgements}

We thank Tom Hartman, Diego Hofman, and especially Clay C\'ordova for helpful comments and conversations.  DP also 
thanks the Aspen Center for Physics for its hospitality during the completion of this work.  This work is supported in part by the 
Harvard Center for the Fundamental Laws of Nature and by NSF grant PHY-0556111.

\appendix

\section{Conventions}
\label{app:conventions}

Our metric and spinor conventions are those of the $\eta_{ab}=\mathrm{diag}(-1,+1,+1,+1)$ version of~\cite{Dreiner:2008tw}.  The Clifford relation is $\s_a\bar\s_b+\s_b\bar\s_a=-2\eta_{ab}$, so that one can convert between vectors and bispinors as $(x)_{\a\dot\a}=x_a\s^a_{\a\dot\a}$ and $x^a=-\frac 1 2\tr(\bar\s^a x)$.  These conventions agree with those of Wess and Bagger~\cite{Wess:1992cp} and Osborn~\cite{Osborn:1998qu}, with a single exception --- the sign of $\s^0$, which affects the coefficient of $\e^{abcd}$ in products of $\s$'s and $\bar \s$'s.  Specifically, we have
\ben
\s^a\bar\s^b\s^c &=& -\eta^{ab}\s^c+\eta^{ca}\s^b-\eta^{bc}\s^a-i\e^{abcd}\s_d\qquad(\textrm{this paper})\\
\s^a\bar\s^b\s^c &=& -\eta^{ab}\s^c+\eta^{ca}\s^b-\eta^{bc}\s^a+i\e^{abcd}\s_d\qquad(\textrm{W\&B})
\een
To convert between these conventions, one simply flips the sign of $\e^{abcd}$ wherever it appears.

For the $\cN=1$ superconformal algebra $\SU(2,2|1)$, we follow the conventions used in~\cite{Butter:2009cp}; in particular we take bosonic generators to be {\it anti-hermitian} (that is, they differ from the usual definitions by a factor of $i$). This eliminates some factors of $i$ from the commutation relations, somewhat simplifying the algebra in Section~\ref{sec:scftblocks}.

Let us arrange the superconformal generators according to their dimensions and spins as follows
\ben
\label{eq:algebratable}
\begin{array}{cccccc}
\dim(X)\quad &&&&&\\
+1\quad& & & P_{a} & &\\
+1/2\quad&& Q_{\a} & & \bar Q_{\dot\a} &\\
0\quad&M_{\a\b} & & D,R & & M_{\dot\a\dot\b}\\
-1/2\quad& & S_\a  & & \bar S_{\dot\a} &\\
-1\quad& & & K_a, & &
\end{array}
\een
where $M_{\a\b}=(\s^{ba}\e)_{\a\b}M_{ab}$ and $M^{\dot\a\dot\b}=(\bar \s^{ba}\e)^{\dot\a\dot\b}M_{ab}$ are self-dual and anti-self-dual rotation generators.

The dilatation operator and $U(1)_R$ generator act as
\begin{align}
[D,X]&=\dim(X)X & [R,X]&=i\,r(X)X,
\end{align}
where $X$ is any generator, $\dim(X)$ is given in the above table~(\ref{eq:algebratable}), and $r(X)$ is the $R$-charge of $X$, given by $+1$ for $X=S,\bar Q$, by $-1$ for $X=Q,\bar S$, and zero otherwise.  The additional commutation relations of the conformal sub-algebra are given by
\ben
\,[M_{ab},P_c ] &=& P_a \eta_{bc} - P_b\eta_{ac}, \qquad [ M_{ab},K_c ] \,\,\,=\,\,\, K_a \eta_{bc} - K_b\eta_{ac}\nonumber\\
\,[M_{ab},M_{cd}] &=& \eta_{bc}M_{ad}-\eta_{ac}M_{bd}-\eta_{bd}M_{ac}+\eta_{ad}M_{bc}\nonumber\\
\,[K_a,P_b] &=& 2\eta_{ab} D-2 M_{ab}.
\een
Rotation generators act on spinors as
\begin{align}
[M_{ab},X_\a]&=(\s_{ab})_\a{}^\b X_\b & [M_{ab},\bar X^{\dot\a}]&=(\bar\s_{ab})^{\dot\a}{}_{\dot\b}\bar X^{\dot\b},
\end{align}
where $X_\a=S,Q$ and $\bar X=\bar Q,\bar S$.  Finally, the remaining non-vanishing commutation relations involving fermionic generators are
\begin{align}
\{Q_\a,\bar Q_{\dot\a}\} &= -2i\s_{\a\dot\a}^a P_a,&
\{S_\a,\bar S_{\dot\a}\} &= +2i\s_{\a\dot\a}^a K_a\\
\ [K_a,Q_\a] &= i\s_{a\a\dot\b} \bar S^{\dot\b},&
\ [S_\a,P_a] &= i\s_{a\a\dot\b}\bar Q^{\dot\b}\\
\ [K_a,\bar Q^{\dot\a}] &= i\bar \s_a^{\dot\a\b}S_\b, &
\ [\bar S^{\dot\a},P_a] &= i\bar\s_a^{\dot\a \b}Q_\b,
\end{align}
\begin{align}
&\{S_\a,Q_\b\} = 2D\e_{\a\b}-2 M_{\a\b}-3i R \e_{\a\b},\\
&\{\bar S^{\dot\a},\bar Q^{\dot\b}\} = 2D\e^{\dot\a\dot\b}-2 M^{\dot\a\dot\b}+3i R\e^{\dot\a\dot\b}.
\end{align}

The relation between our conventions for the super-Poincar\'e subalgebra, and those of Wess and Bagger is summarized by equating supergroup elements at each point $(x,\th,\bar\th)$ in superspace
\ben
\lr[{e^{x\.P+\th Q+\bar\th \bar Q}}]_\textrm{this paper} &=& \lr[{e^{i(-x\.P+\th Q+\bar\th \bar Q)}}]_\textrm{W\&B}.
\een
In particular, component expansions of our superfields $\cO(x,\th,\bar\th)=e^{x\.P+\th Q+\bar\th\bar Q}\cO(0)$ are the same as component expansions in Wess and Bagger, with the only difference being an overall factor of $i$ or $-i$ in the action of super-Poincar\'e generators.

\section{Implementation Details}
\label{app:implementation}

In this Appendix, we discuss some details of our implementation of linear programs for extracting bounds from crossing relations.  We first manipulate the crossing relation into a useful form, and then discuss efficient methods for calculation. Finally, we summarize our choice of programs and parameters for generating the bounds in this paper.

\subsection{Explicit Formulae for Linear Functionals}

Using the explicit expression (\ref{eq:explicitconformalblocks}), we can rewrite the crossing relations Eq.~(\ref{eq:crossing}) and (\ref{eq:crossingphiphis}) as
\ben
\label{eq:crossingrelationnice}
\left[\frac{(z-\bar{z})}{[(1-z)(1-\bar{z})]^d} - \frac{(z-\bar{z})}{(z\bar{z})^d}\right] &=& \sum_{\De,l} \frac{|\l_\cO|^2}{2^l} \left[ \frac{k_{\De+l}(z) k_{\De-l-2}(\bar{z})}{(z\bar{z})^{d-1}} + \frac{k_{\De+l}(1-z) k_{\De-l-2}(1-\bar{z})}{[(1-z)(1-\bar{z})]^{d-1}}\right]\nonumber\\
&&\qquad\qquad- (z \leftrightarrow \bar{z}),
\een
where the left-hand side is the contribution of the unit operator, and the sum is over the appropriate spectrum of primaries appearing in the OPE.  Note that in the charged-scalar case, the $(-1)^l$ factor in the conformal blocks cancels with the $(-1)^l$ in Eq.~(\ref{eq:crossingphiphis}), so that odd-spin contributions are not qualitatively different from even-spin contributions.

We could bring this into the form (\ref{eq:crossingrewrite}) by additionally dividing by the left-hand side and isolating the term in the sum corresponding to a particular operator $\cO_0$.  We would then consider the space of linear functionals $\a:f(z,\bar{z})\mto \sum_{m+n\leq 2k}a_{mn}\ptl_z^m\ptl_{\bar z}^n f(1/2,1/2)$ applied to both sides.  Note however that we get the same space of functionals if we do not first divide by the unit operator, since derivatives of a product are linear combinations of derivatives of the two factors.  Thus, to implement the algorithm described in Section~\ref{sec:boundsfromcrossing}, we can simply compute derivatives at $z=\bar z=1/2$ of Eq.~(\ref{eq:crossingrelationnice}) as written.

Because of symmetry under $(z,\bar z)\leftrightarrow(1-z,1-\bar z)$ and antisymmetry under $z\leftrightarrow \bar z$, derivatives $\ptl_z^m\ptl_{\bar z}^n$ at $(1/2,1/2)$ will vanish unless $m\neq n$ and $m+n$ is even.  Further, it suffices to take $m<n$ by symmetry.  Thus, our $\cW_k$ are precisely defined as the space of real linear functionals
\ben
\label{eq:linearfunctionalspace}
\a:f(z,\bar z)\mto \sum_{\stackrel{m+n\leq 2k}{m+n\in 2\Z,m<n}}a_{mn}\ptl^m_z\ptl^n_{\bar z}f(1/2,1/2),
\een
which has dimension $\frac{k(k+1)}{2}$.  We will write the coefficients $a_{mn}$ collectively as a vector $\ba$.

From Eq.~(\ref{eq:crossingrelationnice}), we see that the building blocks of these functionals are derivatives of $z^{1-d}k_\b(z)$ at $z=1/2$.  These have an analytic expression in terms of hypergeometric functions which we can derive by matching power series,
\ben
\sum_{n=0}^{\oo} C^{n}_{\beta,d} \frac{(z-1/2)^n}{n!}
&\equiv& z^{1-d+\b/2}{}_2F_1(\b/2,\b/2,\b,z)\nonumber\\
&=& 
\sum_{m=0}^{\oo} \frac{\Gamma(\beta) \Gamma(\beta/2 +m)^2}{\Gamma(\beta/2)^2 \Gamma(\beta+m)} \frac{z^{m+1-d+\b/2}}{m!} \nonumber\\
&=& \sum_{n,m=0}^{\oo} \frac{\Gamma(\beta) \Gamma(\beta/2 +m)^2 \Gamma(m+\beta/2-(d-1)+1)}{\Gamma(\beta/2)^2 \Gamma(\beta+m) \Gamma(m+\beta/2-(d-1)+1-n)}\nonumber\\
&&\qquad\x\left(\frac12\right)^{m-n+\beta/2-(d-1)} \frac{(z-1/2)^n}{n! m!}.
\een
Performing the $m$-summation finally determines the coefficients
\ben
\label{eq:Ccoefficients}
C^{n}_{\beta,d} &=& 2^{n+(d-1)-\beta/2} \frac{\Gamma(\beta/2+2-d)}{\Gamma(\beta/2+2-d-n)} {}_3F_2(\beta/2+2-d,\beta/2,\beta/2;\beta/2+2-d-n,\beta;1/2).\nonumber\\
\een

Now $\ptl_z^m\ptl_{\bar z}^n|_{(1/2,1/2)}$ applied to Eq.~(\ref{eq:crossingrelationnice}) can be written
\ben
\label{eq:derivsofcrossing}
\hspace{-.3in}
\frac{2^{2d+n+m-1}(d-1)(n-m)\Gamma(1-d)^2}{\Gamma(2-d-n)\Gamma(2-d-m)} &=& \sum_{\De,l} \frac{|\l_\cO|^2}{2^l} \lr[{C^{n}_{\De+l,d} C^{m}_{\De-l-2,d} - C^{m}_{\De+l,d} C^{n}_{\De-l-2,d}}]
\!,
\een
where we have assumed that $m+n$ is even.  Hence the objective function ``$\a(1)$" in our linear program is given by $\ba\mto \bv\.\ba$, where $\bv$ is a vector of values of the left-hand side of Eq.~(\ref{eq:derivsofcrossing}) for different $m$ and $n$ (depending on our choice of $\cW_k$).  Meanwhile, each constraint $\a(F_{\De,l})\geq 0$ becomes $\bu_{\De,l}\.\ba\geq 0$, where $\bu_{\De,l}$ is a vector of values of
\ben
\frac{1}{2^l} \left[C^{n}_{\De+l,d} C^{m}_{\De-l-2,d} - C^{m}_{\De+l,d} C^{n}_{\De-l-2,d}\right]
\een
for different $m,n$.

\subsection{Optimizations}

Before running each linear program, we must compile a list of $\bu_{\De_i,l_i}$ for all $(\De_i,l_i)$ in our choice of discretization $D$.  We found in practice that simply evaluating the expression~(\ref{eq:Ccoefficients}) for $C_{\b,d}^m$ in {\tt Mathematica} introduced a performance bottleneck.  One possible remedy is to precompute values of $C^m_{\b,d}$ for different $m,\b,d$, and then perform table lookups to compile the list of constraints $\bu_{\De_i,l_i}$.  However, because there are three parameters $m,\b,d$ to scan over, this would require a lot of memory and some careful bookkeeping.

An alternative approach uses the fact that $z^{1-d}k_\b(z)\equiv u_{\b,d}(z)$ satisfies a simple differential equation which implies a recursion relation for its derivatives.  Using the hypergeometric differential equation for ${}_2F_1(\b/2,\b/2,\b,z)$, and conjugating the resulting differential operator by $z^{\b/2+1-d}$, we find
\ben
0 &=& \p{z^2(1-z)\frac{\ptl^2}{\ptl z^2}+z(2d(1-z)+z-2)\frac{\ptl}{\ptl z}+(d-1)(d(1-z)+z-2)-\l_\b}u_{\b,d}(z),\nonumber\\
\een
where $\l_\b\equiv \frac 1 4\b(\b-2)$.  Now taking $n-2$ derivatives with respect to $z$ and evaluating at $z=1/2$, we find the recursion relation
\ben
\label{eq:coeffrecursion}
C^n_{\b,d} &=& 2(5-2d-n)C^{n-1}_{\b,d}+4\p{2\l_\b+n(n-3)-d^2+4d-1}C^{n-2}_{\b,d}\nonumber\\
&&+\,8(n-2)(n+d-4)^2 C^{n-3}_{\b,d}.
\een
This can be iterated to give
\ben
\label{eq:recursionsolution}
C^n_{\b,d} &=& P_n(\b,d)2^{d-1}k_\b(1/2) + Q_n(\b,d)2^{d-1} k'_\b(1/2),
\een
where $P_n$ and $Q_n$ are polynomials in $\b$ and $d$.  Now Eq.~(\ref{eq:recursionsolution}) can be made extremely computationally efficient.  We first determine $P_n$ and $Q_n$ for all $0\leq n\leq 2k$ using (\ref{eq:coeffrecursion}).  Additionally, we precompute a table of $k_\b(1/2)$ and $k_\b'(1/2)$ for different $\b$ values.  This reduces the evaluation of $C^n_{\b,d}$ to simple polynomial and exponential evaluation, along with two table lookups.

\subsection{Programs and Parameters}

Here, we give an account of the programs and parameters used to generate the bounds in Section~\ref{sec:bounds}.  In each linear program, we take a discretization of the form
\ben
\label{eq:discretization2}
D &=& \{(\De_\mathrm{min}+n\e,l):n=0,\dots,N\textrm{ and }l=0,2,\dots,L\},
\een
where $\De_\mathrm{min}$ depends on the problem at hand, as discussed in the text.  In addition to the parameters $\e,N,$ and $L$, one must also pick a subspace $\cW_k\subset \cV^*$.  Our choices in this paper, along with the resolution of our plots are as follows:

\begin{center}
\begin{tabular}{c | l | c | c | c | c | c}
Fig. & Title & $N\e$ & $L$ & $\e$ & $k$ & Resolution\\
\hline
\ref{fig:dimensionbound} & $\max(\De_{\Phi^\dag\Phi})$ & 50 & 25 & 0.02 & 6 & $\de d = 0.0025$\\
\ref{fig:SUSYscalarOPE} & $\max|\l_{\cO_0}|\text{ for }l_0=0$ & 30 & 25 & 0.02 & 5 & $\de\De_0 = 0.025$\\
\ref{fig:chargedFlavorPlot} & $\max(\tau_{IJ}T^I T^J)\text{ for a charged scalar}$ & 20 & 25 & 0.05 & 5 & $\de d=0.05$\\
\ref{fig:SUSYFlavorPlot} & $\max(\tau_{IJ}T^I T^J)\text{ for a chiral primary}$ & 20 & 25 & 0.05 & 5 & $\de d=0.05$\\
\ref{fig:realcentralcharge} & $\min(c)$ for a general CFT & 80 & 25 & 0.02 & 3,4,5,6 & $\de d=0.01$\\
\ref{fig:SUSYcentralchargeplot} & $\min(c)$ for a SCFT & 20 & 25 & 0.05 & 5 & $\de d=0.01$
\end{tabular}
\end{center}

We have chosen $N\e$ and $L$ large enough so that the optimal linear combination satisfies $\a_*(F_{\De,l})\geq 0$ asymptotically as $\De,l\to\oo$.  At any finite $\e>0$, one can expect violations of the constraints $\a(F_{\De,l})\geq 0$ of order $\e^2$ at isolated $\De,l$.  Decreasing $\e$ reduces these effects, but has a computational cost since the linear programming algorithm we use (the simplex algorithm) runs in $O(1/\e^3)$ time (cubic in the number of constraints).  In each case above, we have verified that changing $\e$ slightly does not appreciably affect the results, so that we believe our plots accurately reflect the $\e\to 0$ limit.  The curves themselves were generated by computing points with the resolution specified above (dropping a small number of points where the linear program was not well-behaved) and plotting an interpolating function.  

We generated the input data for each linear program with {\tt Mathematica}.
For actually solving linear programs, we used the GNU Linear Programming Kit ({\tt glpk}) \cite{glpk}, which seemed generally faster and less unpredictable than {\tt Mathematica}'s {\tt LinearProgramming} routine.  Most of our computations were run on Harvard's Odyssey cluster supported by the FAS Sciences Division Research Computing Group.



\end{document}